\documentclass[11pt]{article}
\usepackage[utf8]{inputenc}
\usepackage[margin=0.8in]{geometry}
\usepackage{graphicx}

\usepackage{comment,subfigure,color}
\usepackage{comment}
\usepackage{booktabs} 
\usepackage{array} 
\usepackage{paralist} 
\usepackage{verbatim} 
\usepackage{subfig} 
\usepackage[stable]{footmisc}
\usepackage{wrapfig}
\usepackage[font=footnotesize,labelfont=bf]{caption}
\usepackage{fancyhdr} 
\usepackage{color}
\usepackage{cite}
\usepackage{sectsty}
\allsectionsfont{\sffamily\mdseries\upshape} 

\usepackage[nottoc,notlof,notlot]{tocbibind} 
\usepackage[titles,subfigure]{tocloft} 
\usepackage{titlesec}
\titleformat{\section}[block]{\fontfamily{cmr}\bfseries\filcenter}{\thesection}{1em}{}
\titleformat{\subsection}
  {\fontfamily{cmr}\bfseries\upshape}
  {\thesubsection}
  {1em}{}
\titlespacing\section{0pt}{5pt plus 2pt minus 2pt}{0pt plus 2pt minus 2pt}
\titlespacing\subsection{0pt}{5pt plus 2pt minus 2pt}{0pt plus 2pt minus 2pt}

\makeatletter
\renewcommand{\fnum@figure}{Fig. \thefigure}
\makeatother
\newcommand{\bey}{\begin{eqnarray}}
\newcommand{\eey}{\end{eqnarray}}
\newcommand{\vep}{\varepsilon}
\newcommand{\fvep}{\fg \varepsilon}
\newcommand{\sg}{\sigma}
\newcommand{\fsg}{\fg \sigma}
\newcommand{\bec}{\begin{center}}
\newcommand{\eec}{\end{center}}
\newcommand{\pmb}[1]{{\setbox0=\hbox{#1}%
  \kern-.025em\copy0\kern-\wd0
  \kern.05em\copy0\kern-\wd0
  \kern-.025em\raise.0433em\box0 }}
\newcommand{\fg}[1]{\mbox{\pmb{$#1$}}}

\begin{document}
\bec
{\bf \large  Resolving puzzles of the phase-transformation-based mechanism of the deep-focus earthquake }
\\
Valery I. Levitas\\
{\it
Iowa State University, Departments of Aerospace Engineering and
Mechanical Engineering,  Ames, Iowa 50011, USA\\
Ames Laboratory,
Division of Materials Science and Engineering,  Ames, IA, USA}

\eec

\begin{figure}[ht]
\vspace{-3mm}
\centering
\includegraphics[width=0.8 \linewidth]{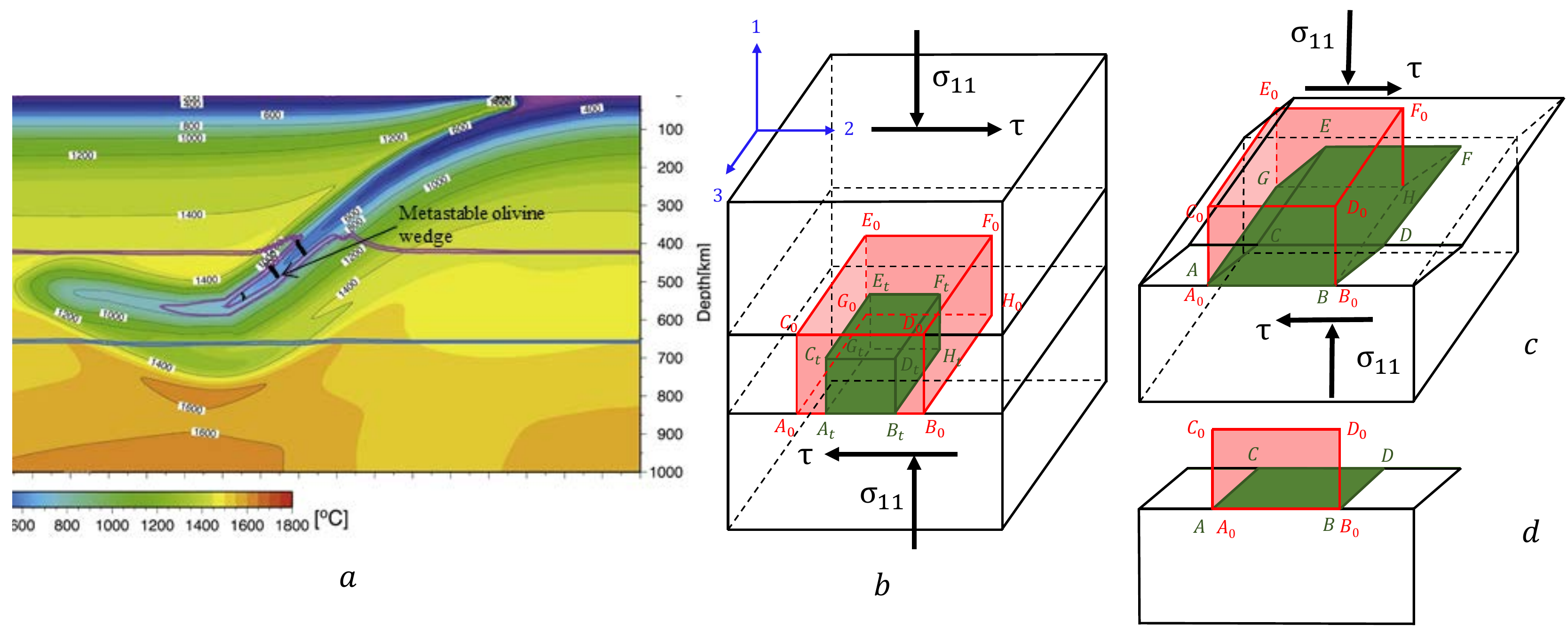}
\caption{{\bf Schematics of triggering deep focus earthquake by  transformation-deformation-heating bands during
PT from the subducted metastable olivine  to spinel.} (a) Results of modeling of subduction of the Pacific plate including metastable olivine wedge beneath Japan with the temperature contour line. Magenta lines denote 1\% (upper line) and 99\% (lower line) of PT from olivine to $\beta$-spinel; blue lines denote 1\% (upper line) and 99\% (lower line) of PT from $\gamma$-spinel to bridgmanite+magnesiowüstite.
Black lines designate
transformation-deformation-heating bands. Earthquakes occur at the olivine wedge boundary   (adapted with modifications from \cite{Kawakatsu-11} with permission from Elsevier Publ.). (b) Schematics of a  transformation-deformation-heating band within a rigid  space. Part of a band before PT (red) and after PT and isotropic transformation strain (green) is shown. (c)
To satisfy the continuity of displacements across the shear-band boundary and rigid space outside the band, additional TRIP
 develops, leading  to deformation of the green rectangular $A_tB_tG_tH_t$ to $ABGH$ that coincides with $A_0B_0G_0H_0$ and to   large plastic shear. (d) 2D view (along axis 3) of (c). }
\label{Fig-shear-band}
\vspace{-2mm}
\end{figure}

Deep-focus earthquakes that occur at 350–660 km, where  pressures $p=$12-23 GPa   and  temperature $T=$1800-2000 K, are
generally assumed to be caused by olivine$\rightarrow$spinel phase transformation, see pioneering works \cite{Kirby-87,Burnley-Green-89,green-etal-15,green-burnley-89,Kirbyetal-96,green-etal-Science-13,Wangetal-17,Green-07,Green-17,Zhan-20}.
However, there are many existing puzzles:
(a) What are the mechanisms for jump from geological  $10^{-17}-10^{-15}\, s^{-1}$ to seismic $10-10^3\, s^{-1}$ (see \cite{green-etal-15}) strain rates? Is it possible without phase transformation?
(b) How does metastable olivine, which  does not completely transform to spinel at high  temperature and  deeply in the region of stability of spinel for over the million years, suddenly transforms during seconds and
generates seismic strain rates $10-10^3\, s^{-1}$  that   produce strong seismic waves?
(c) How to connect deviatorically dominated seismic signals with volume-change dominated transformation strain during phase transformations  \cite{Frohlich-89,Zhan-20}?
Here we introduce a combination of several novel concepts that allow us to resolve the above puzzles quantitatively.
We treat the transformation in olivine like plastic strain-induced (instead of pressure/stress-induced)
and find  an analytical 3D solution for   coupled deformation-transformation-heating processes
in a shear band. This solution predicts conditions for severe (singular) transformation-induced plasticity (TRIP) and
self-blown-up deformation-transformation-heating process    due to positive thermomechanochemical feedback
between TRIP and  strain-induced transformation. In nature, this process leads to  temperature in a band exceeding the unstable stationary temperature, above which the self-blown-up shear-heating
process in the shear band occurs after  finishing the phase transformation. Without phase transformation and TRIP, significant temperature and strain rate increase is impossible. Due to the much smaller band thickness in the laboratory,
heating within the band does not occur, and plastic flow after the transformation is very limited.
Our findings change the main concepts in studying the initiation of the deep-focus earthquakes and phase transformations during plastic flow in geophysics in general. The latter may change the interpretation of different geological phenomena, e.g., the possibility of the appearance of microdiamond directly in the cold Earth crust within shear bands  \cite{Gaoetal-17} during tectonic activities without subduction to the mantle and uplifting. Developed theory of the self-blown-up transformation-TRIP-heating process is applicable outside geophysics for various processes in materials under pressure and shear, e.g., for new routes of material synthesis  \cite{Gaoetal-17,Levitas-MT-19}, friction and wear, surface treatment, penetration of the projectiles and meteorites, and severe plastic deformation and mechanochemical technologies.
\\
\\
Deep-focus earthquakes  are  very old puzzles in geophysics.
While  the shallow earthquakes occur due to brittle fracture, materials at 350–600 km are
above the brittle–ductile transition \cite{Frohlich-89}. That is why the  main hypothesis is that
the earthquakes are caused by instability due to phase transformation (PT)  from the subducted metastable $\alpha$-olivine (forsterite) to denser  $\beta$-spinel (wadsleyite) or $\gamma$-spinel (ringwoodite) \cite{Kirby-87,Burnley-Green-89,green-etal-15,green-burnley-89,Kirbyetal-96,green-etal-Science-13,Wangetal-17,Green-07,Green-17,Zhan-20}
(Figs. \ref{Fig-shear-band}a and S7).
Self-organized ellipsoidal transformed regions (anticracks)  filled with nanograined product phase with very low shear resistance and orthogonal to the largest normal stress were considered. A set of anticracks aligned along the maximum shear stress reduces shear resistance and causes a shear band. In \cite{Meade-Jeanloz-89,Meade-Jeanloz-91}, the acoustic emission approach was pioneered to detect "seismic" events during several PTs, which was interpreted in favor of PT and shear instability hypotheses of the earthquake initiation. However, these semi-qualitative approaches cannot resolve puzzles mentioned in the abstract.
\\\\
{\bf Plastic strain-induced phase transformations}
\\
It is clear that to obtain such jumps in plastic flow and PT rates in some rare cases, theory should contain singularity that strongly depends on some external conditions. To resolve the problem, we will utilize the main concept of high-pressure mechanochemistry
\cite{levitas-mechchem-04,levitas-prb-04,Levitas-JPCM-18,Levitas-MT-19}.
Our first point is that
 in  all previous geophysical papers \cite{Kirby-87,Burnley-Green-89,green-etal-15,green-burnley-89,Kirbyetal-96,green-etal-Science-13,Wangetal-17,Green-07,Green-17,Zhan-20}, pressure- and stress-induced PTs  were considered a mechanism for initiating the shear instability.
 These PTs start at crystal defects that naturally exist in material and for stresses below
the yield strength. These defects (e.g., various dislocation
structures or grain boundaries) produce stress concentrators
and serve as nucleation sites for a PT. Since a number of such defects is limited, one has to increase pressure to activated defects with smaller stress concentration. In contrast, plastic strain-induced PTs take place by
nucleation at defects produced in the course of plastic flow. The largest   concentration of all stress components can be produced at the tip of the dislocation pileups, proportional to the number of dislocations $N$ in a pileup.   Since $N = 10-100$, local stresses could be huge and exceed the lattice instability limit, leading to nucleation of spinel during $\sim10\, ps$.  Due to a strong reduction of stresses away from the defect tip, growth is very limited. Thus, the next plastic strain increment leading to new defects and new nuclei at their tips is required to continue PT.
That is why time is not a governing parameter in a kinetic equation, and plastic strain plays a role of a time-like parameter (see Eq.(\ref{l-g-4})).
Arrested growth also  explains nanograin structure after strain-induced PTs in various systems \cite{Gaoetal-17,Levitasetal-JCP-BN-06,Cheng-Levitas-etal-12,Blank-Estrin-2014}, including olivine$\rightarrow$spinel
\cite{Riggs-Green-pileup-05,green-etal-15,Green-17,Wangetal-17}. 
The important
point is that the deviatoric (nonhydrostatic) stresses in the nanoregion near the defect tip are not bounded by the engineering yield strength but rather by the ideal strength in shear for a defect-free lattice which may be higher by a factor of 10 to 100. Local
stresses of such magnitude may result in the nucleation of the high-pressure phase at an applied pressure that is not only
significantly lower than that under hydrostatic loading but also below the phase-equilibrium pressure. For example,
plastic strain-induced PT from graphite to hexagonal and cubic diamonds at room temperature was obtained at 0.4 and 0.7 GPa,
 50 and 100 times below than under hydrostatic loading, respectively, and well below the phase equilibrium pressure of 2.45 GPa \cite{Gaoetal-17}
 (see other examples for PTs in Zr, Si, and BN  \cite{Levitas-Shvedov-02,Pandey-Levitas-ActaMat-20,Blank-Estrin-2014,Cheng-Levitas-etal-12}). In addition, such
highly-deviatoric stress states with large stress magnitudes cannot be realized in bulk. Such unique stresses may lead to
PTs into stable or metastable phases that were not or could not be attained in bulk under hydrostatic or quasi-hydrostatic
conditions \cite{Cheng-Levitas-etal-12,levitasetal-SiC-12,Blank-Estrin-2014,Edalati-Horita-16}.
It was concluded in \cite{levitas-mechchem-04,levitas-prb-04,Levitas-JPCM-18,Levitas-MT-19} that plastic strain-induced
transformations require very different thermodynamic,
kinetic, and experimental treatments than pressure- and
stress-induced transformations.

Strain-induced character of PTs is consistent with results in  \cite{Riggs-Green-pileup-05,green-etal-15,Green-17}, where metastable  olivine Mg$_2$GeO$_4$
transforms into spinel in the 70 nm thick shear band, partially transforms in the surrounding band of few $\mu m$ thick, and does not transform away from the band.
These thin planar layers of strain-induced nanograined (10-30 nm) Mg$_2$GeO$_4$  spinel within olivine 
were observed in \cite{Riggs-Green-pileup-05,green-etal-15} after laboratory experiment and suggested as an additional to anticrack mechanism
of shear weakening.
They
appear along the specific slip planes, are related to dislocation pileups, and correspond to our model's prediction below.
The lower temperature is, the more strain-induced planar spinel bands  and less stress-induced spinel anticrack regions are observed,
consistent with promoting effect of strain-induced defects.
Relative slip along a 70 nm thick transformed planar layer is 3 microns, i.e., shear strain $\gamma=43$; slip rate is $1\, \mu m\, s^{-1}$, thus
$\dot{\gamma}=14\, s^{-1}$ and time of sliding (and PT) is $\gamma/\dot{\gamma}=3 \,s$.
These bands offset multiple non-transforming pyroxene crystals, which allows determining relative slip.
In contrast to anticracks that are mostly orthogonal to the compressive stress, transformation bands are mostly under $45^0$ with some scatter
 to the compression direction, i.e., they coincide with planes with  maximum shear stress, or pressure-dependent resolved shear stress.
In nature, the Punchbowl Fault also exhibited
a few-mm thick  slip zone, along
which slip occurred by several kilometers,
which contains nanograins \cite{green-etal-15,Green-17}, i.e., shear strain $\gamma=10^6$.
Similar strain-induced PTs and reactions are observed at the surface layers in friction experiments  \cite{green-etal-15,Green-17}.

\begin{figure}[ht]
\vspace{-3mm}
\centering
\includegraphics[width= \linewidth]{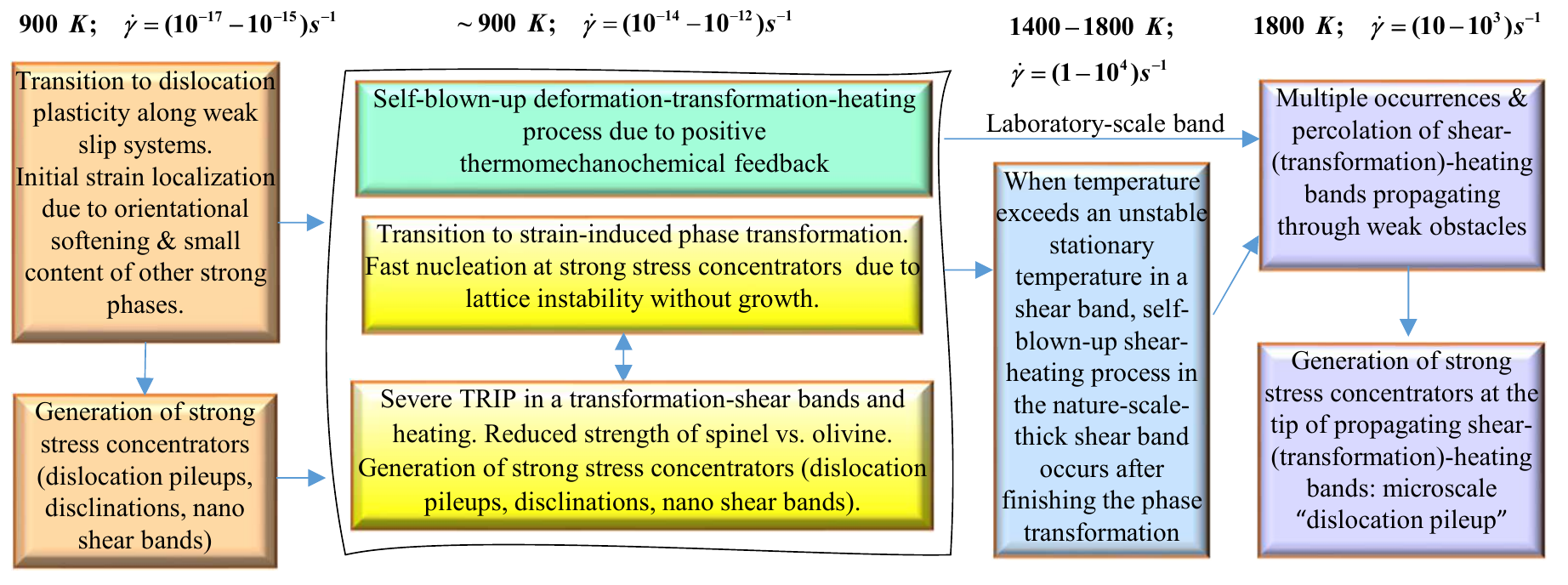}
\caption{{\bf Mechanisms of localized thermoplastic flow and PT leading to high strain and PT rates and high temperatures in a transformation-deformation and shear bands.} Temperature and shear rate before each stage is shown on the top.
Initial strain localization occurs due to transition to dislocation plasticity along properly oriented weak $[001](010)$ slip systems and                      corresponding  orientational softening, as well as along the plane with a small content of other strong phases like diopside.
Localized plastic flow leads to the generation of strong stress concentrators (dislocation pileups, disclinations, shear nanobands), causing strain-induced PT. Due to crystal lattice instability, fast nucleation at strong stress concentrators  occurs during 10 ps but without growth, leading to a weaker nanograined spinel and strain-controlled kinetics proportional to the strain rate instead of time. Volume reduction during PT in a shear band causes severe TRIP,
which in turn causes strain-induced PT leading to further TRIP and PT, and so on. This positive  thermomechanochemical feedback leads to self-blown-up deformation-transformation-heating up to high temperature, exceeding unstable stationary temperature in a shear band, and high strain rate. After completing PT, in nature but not in  the thin laboratory-scale band, further heating and increased strain rate in a band occur due to shear flow. Similar processes are expected in multiple transformation-deformation and then just deformation bands that find ways through weak obstacles and may percolate or just increase the total shear-band volume and amplify generated seismic waves. Propagating transformation-deformation and just plastic shear bands generate  strong stress concentrators at their tips producing  a microscale counterpart of a dislocation pileup, which  causes both fast PT and plasticity and further propagation of a shear band, i.e., repeats the above processes at a larger scale.}
\label{fig:cascade}
\vspace{-2mm}
\end{figure}
The suggested mechanisms of localized thermoplastic flow and PT consist of several interrelated steps shown in Fig. \ref{fig:cascade} and will be elaborated below.
\\\\
{\bf Mechanisms and conditions of localized thermoplastic flow and heating}
\\
According to \cite{Billen-20}, seismicity in the transition zone correlates with the rate of plastic flow, which is in the range of
$10^{-17}-10^{-15}\, s^{-1}$.
Orthorhombic olivine has only three independent slip systems set, i.e.,  less than five required for the accommodation of arbitrary homogeneous deformation. That is why other mechanisms like grain-boundary migration through disclination motion \cite{Cordier-Fressengeas-14}, amorphization \cite{Samae-Cordier-etal-21},  dislocation climb,  diffusive creep, and other isotropic mechanisms with linear flow rule \cite{Hirth-Kohlstedt-95,Raterron-etal-14} supplement dislocation plasticity and control strain rate. Less than 40\% of olivine aggregate strain at high temperatures may be accommodated by   dislocation activity. However, when one of the slip system is aligned along or close to maximum shear stress, faster shear-dominated deformation is possible controlled solely by dislocations. Especially, $[001](010)$ slip system has critical shear stress of 0.15 MPa, at least 3 times lower
than that  for all other systems (at 405 km depth, $T=1757 K$, $p=13.3\, GPa$, equivalent plastic strain rate $\dot\vep=10^{-15}$)  \cite{Raterron-etal-14}. Thus, if some group of grains is oriented with $[001](010)$ slip system along the maximum shear stress,
dislocation glide may occur compatible with shear strain localization due to orientational softening.
Despite the variety of deformation mechanisms, plastic flow in olivine and spinel is formally described by
\bey
\dot{\vep}= H \sg^n \exp(-Q_r/T) \;   \rightarrow \; M= \dot{\vep}(T)/\dot{\vep}(T_0)=\exp[-Q_r(T^{-1}-T_0^{-1})],
\label{Eq-1}
\eey
where $Q_r=Q/R$, 
$Q$ is the activation energy, $R$ is the gas constant, and $\sg$ is the differential stress, which is the same within and outside of the shear band due to continuity of shear stresses along the band boundary.
Since for  olivine $n=3.5$  \cite{Ogava-87,Raterron-etal-14}, reduction in resistance by a factor of 3
leads for the same stress to increase in the strain rate by a factor of 47.
Also, in Earth, olivine is mixed with diopside,  which has much higher critical shear stresses, 7.31-64.7 MPa at the same conditions\cite{Raterron-etal-14}. 
Thus, shear localization should start in the region with small diopside content, which may also increase strain rate by additional one-two orders of magnitude. In total, when both proper alignment of olivine grains and small diopside content are combined, local strain rate may increase by up to $10^3$ times without a change in temperature and reach  $10^{-14}-10^{-12}\, s^{-1}$. At such a strain rate, shear localization may be promoted by plastic heating in a band with the width $h$ exceeding $10$ to $10^3\ m$ \cite{Ogava-87}, but a characteristic time of this localization, $10$ to $10^4$ years, is way too long to resolve puzzles mentioned in abstract, and too broad to reproduce a few-mm thick  slip zone in the Punchbowl Fault
\cite{green-etal-15,Green-17}. Also, such a slow heating increase chances for slow and nonlocalized olivine-spinel PT, which eliminates the
possibility of fast and localized PT and TRIP described in the next section.

In addition, we can include softening due to the substitution of olivine to a weaker nanograined
spinel in a band.
The initial yield strength in compression $\sg_y$ of the transformed nanograined $\gamma$-spinel at $\dot\vep \simeq 10^{-5} s^{-1}$ is 4.7 times lower than that for olivine \cite{Karato-20}.  According to Eq.(\ref{Eq-1}), this leads to an additional  increase is strain rate for the same stress by a factor of $4.7^{3.5}=225$. Assuming that at a higher strain rate increase in stress in slightly higher, $5.8$, we obtain $5.8^{3.5}=470$. However, we have to exclude a factor of 47 due to orientational softening of olivine (because a factor of 4.7 is in comparison with  normal olivine), leading to an additional increase by a factor of $10$ and strain rate of $10^{-13}-11^{-10}\, s^{-1}$.
Thus, in contrast to \cite{green-etal-15,Green-17}, weak nanograined spinel, while significantly reducing resistance
to shear, increases strain rate less than by a factor of $10^3$.
Anticracks filled with weaker nanograined spinel along the path of a shear band also reduce strength (the main softening mechanism suggested in  \cite{Burnley-Green-89,green-etal-15,green-burnley-89,Green-17}), but much less than   the above estimate when nanograined spinel is located within the entire shear band; that it why we will not consider them.

\begin{figure}[ht]
\vspace{-3mm}
\centering
\includegraphics[width= 0.7\linewidth]{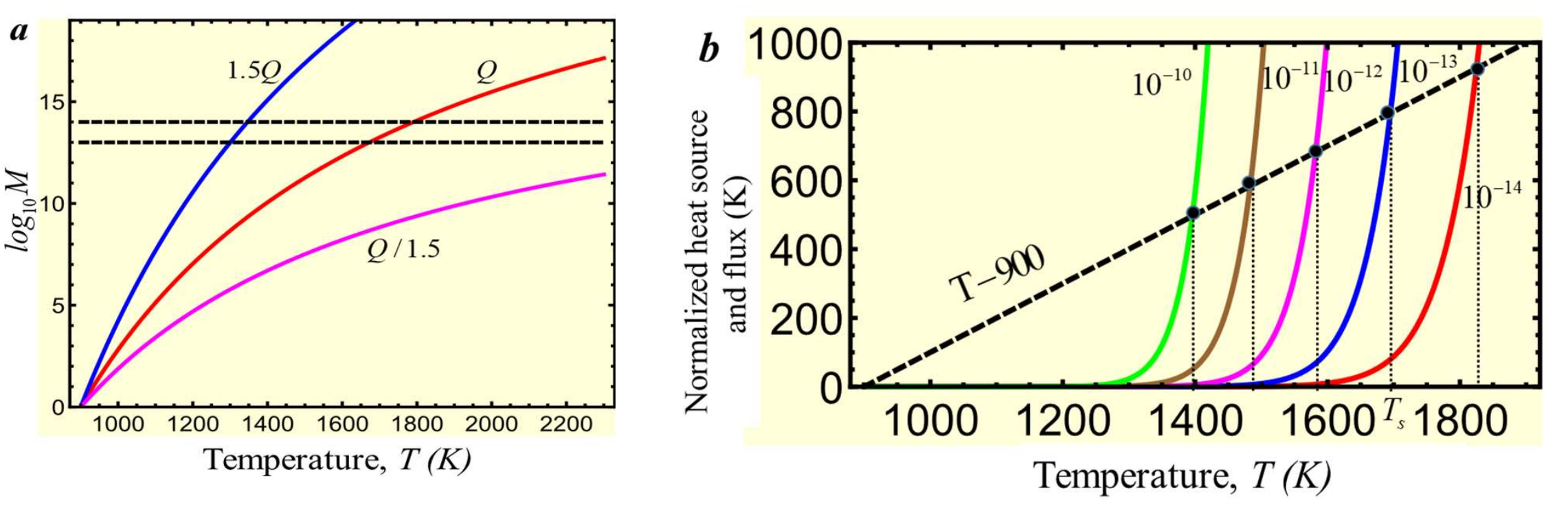}
\caption{{\bf Characteristics of the localized thermoplastic flow.} (a) Plot of $\log_{10}M$ vs. temperature (Eq.(\ref{Eq-1})) for the chosen activation energy $Q/R=58,333 K$ \cite{Ogava-87} and $1.5 Q$ and $Q/1.5$. (b)
Plots of both sides of Eq.(\ref{earth-10}) for stationary temperature, namely the straight line  related to the heat flux from the band and the term related to the plastic dissipation, for different strain rates $\dot{\vep}(T_0)$ (shown near the curves).  Intersections of these lines produce two stationary solutions for the temperature evolution equation. The solution with $T \simeq T_0$ is stable. The second solution $T_s\gg T_0$  is unstable since any fluctuational increase (decrease) in temperature within a band leads to higher (lower) plastic dissipation than the heat flux from the band and further increase (decrease) in temperature. This means that (i) some very significant additional heating source than the traditional plastic flow is required to reach   $T_s$; otherwise, the temperature will be close to $T_0$; (ii) after reaching  $T_s$, plastic dissipation will lead to unlimited heating
up to melting temperature with a corresponding drastic increase in the strain rate. }
\label{fig:temp}
\vspace{-2mm}
\end{figure}

We assume that the initial temperature of the cold slab is $T_0=900K$ \cite{Karato-20}, cold enough to avoid stress-induced olivine-spinel PT in a bulk,  and show that to get the desired jump
in the strain rate, the final temperature should be  $T=1800 K$.
Indeed, taking from \cite{Ogava-87} $Q_r=58,333 K$ we obtain from
Eq.(\ref{Eq-1}) that at $T=1800 K$ the strain rate increases by a factor of $M=10^{14}$ (Fig. \ref{fig:temp}(a)).
Thus, if initial strain rate  in the localized region was $\dot{\vep}(T_0)=10^{-13}-10^{-11}\, s^{-1}$, then after heating to  $T=1800 K$   it increases to $\dot{\vep}(T)=10-10^{3}\, s^{-1}$.
These numbers are close to strain rates of $1-10 \,s^{-1} $ for $\gamma$-spinel at 17 GPa,  1800 K, and grain size of $10 nm$
that can be estimated from Fig. S10 in \cite{Karato-20}.

The temperature evolution equation in a localized shear band with the thickness $h$ and temperature $T$ within the rest of the material with temperature $T_0$ is
\bey
\rho \nu  \dot{T}h=-4 k (T-T_0)/h+ \sg_y \dot{\vep}h = -4 k (T-T_0)/h+   H \sg^{n+1} \exp(-Q_r/T)h
,
\label{earth-9}
\eey
 where $\rho$ is the mass density, $\nu$ is the specific heat, and $k$ is the thermal conductivity.
The term $-4 k (T-T_0)/h$ is the heat flux through two shear-band surfaces due to temperature gradient $2  (T-T_0)/h$, similar to
\cite{Ogava-87}, and  Eq.(\ref{Eq-1})  was used to calculate plastic dissipation. The thermal conductivity $k=\rho \nu \kappa= 2.4 \times 10^{-6} MPa \, m^2/(s \, K)$  \cite{Ogava-87}, where $\kappa= 10^{-6}m^2/s$ is the thermal diffusivity, $\rho= 3000\, kg/m^3$, and $\nu= 800 \, J/(kg\, K)= 800 \times 10^{-6} MPa \, m^3 /(kg\, K)$. Constant $H$ is determined from Eq.(\ref{Eq-1})
as $H=\dot{\vep}(T_0)\sg^{-n}\exp[Q_r/T_0]$. Then the stationary solution $T_s$ of Eq.(\ref{earth-9}) (i.e.,  $\dot{T}=0$) is determined from
\bey
T_s-T_0= 0.25 h^2  \sg \dot{\vep}(T_0)  \exp[-Q_r(T_s^{-1}-T_0^{-1})]/ k.
\label{earth-10}
\eey
Since  the Punchbowl Fault exhibited
a few-mm thick  slip zone \cite{green-etal-15,Green-17}, we assume $h=4 \times 10^{-3}m$.
We also choose  $\sg=300 \, MPa$ \cite{Ogava-87,Karato-20}.

Plots of both sides of Eq.(\ref{earth-10}) in Fig. \ref{fig:temp}(b) shows that there are two stationary solutions. One of the solutions with $T \simeq T_0$ is stable since any fluctuational increase (decrease) in temperature within a band leads to higher (lower) heat flux from the band than the plastic dissipation. The second solution $T_s\gg T_0$ varies from $1396$ to $1825\, K$ when strain rate $\dot{\vep}(T_0)$ reduces from
$10^{-10}$ to $10^{-14}\, s^{-1}$. The higher combination $ h^2  \sg \dot{\vep}(T_0)$ is, the lower the stationary temperature $T_s$ is. This solution  is unstable since any fluctuational increase (decrease) in temperature within a band leads to higher (lower) plastic dissipation than the heat flux from the band and further increase (decrease) in temperature. This means that (a)
localized increase in strain rate and temperature in a thin band is impossible, and  temperature  increase estimated with neglected heat flux term to justify melting \cite{Kanamori-etal-98} or low shear resistance \cite{green-etal-15,Green-17} are wrong (see Supplementary Information); (b)
some very significant additional heating source than the traditional plastic flow is required to reach   $T_s$; otherwise the  temperature will be close to $T_0$; (c) after reaching  $T_s\gg T_0$, plastic dissipation will lead to unlimited heating
up to melting temperature with a corresponding drastic increase in the strain rate. Thus, even if the entire olivine would transform everywhere
to much weaker nanograined spinel (not just in selected anticracks) and softening due to small content of other strong phases (which were not included in the previous models) are present, still, strain rate cannot exceed $\dot{\vep}(T_0)=10^{-13}-10^{-11}\, s^{-1}$, which cannot cause a localized temperature increase.

Note that the transformation heat for olivine-spinel PT increases temperature by $100K$ only \cite{Sung-Burns-76}, which is too small to reach $T_s\gg T_0$.
Below, we suggest PT- and TRIP-related mechanisms of increase in temperature above  $T_s$.
\\\\
{\bf Plastic strain-induced phase transformation olivine$\rightarrow$spinel and TRIP}
\\
Usually,  during a PT,  spinel appears as a continuous film
along grain boundaries with increasing thickness \cite{Mohiuddin-Karato-18} or as anticrack region nucleated at the grain boundaries \cite{Riggs-Green-pileup-05,green-etal-15,Green-17}. Transition to dislocation plasticity should lead to dislocation pileups and strain-induced PT within grains, which is consistent with  band-shaped spinel regions  observed within grains in
\cite{Riggs-Green-pileup-05,green-etal-15,Green-17} and related to dislocation pileups.
Large overdrive and nonhydrostatic stresses promote martensitic PT at dislocations within grains \cite{Poirier-00,Smyth-etal-12}.
Shear stresses at the tip of the dislocation pileup should also change a slow reconstructive mechanism of olivine-spinel PT to a fast martensitic mechanism.  Transformation bands include $(010)$ planes, which include $[001](010)$ slip system with the smallest critical stress,
see \cite{Raterron-etal-14}, consistent with our assumption above. However, there are also  $(011)$ transformation bands, which do not have smaller critical shear stress and do not lead to the orientational softening. That means that orientational softening is not  a mandatory
mechanism for initial localization and can be compensated by smaller diopside content along those planes.

 Strain-controlled kinetic equation \cite{levitas-mechchem-04,levitas-prb-04} for the volume fraction of the strain-induced high-pressure phase simplified in Supplementary Information is
\bey
\frac{d c}{d \vep} = A {\left(1-c \right)}  \qquad  {\rm for} \; p>p_\vep^d (T); \quad A:=
a \frac{p - p_\vep^d (T)}{p_h^d (T) - p_\vep^d (T)} \quad \rightarrow \quad   c=1- \exp(-A \vep).
\label{l-g-4}
\eey
Here, 
$p_{\vep}^d (T)$ and $p_h^d (T)$
  are the minimum pressure at which the direct (i.e., to high-pressure phase) strain-induced and pressure-induced PTs are possible, respectively,
  and $a$ is a  parameter.
 We do not consider strain-induced reverse spinel$\rightarrow$olivine PT, because resultant nanograin spinel deforms dominantly by grain-boundary sliding, which does not produce stress concentrators inside the grains.
  The first experimental and only existing confirmation of  Eq.(\ref{l-g-4})  and parameter identification were   performed  for $\alpha\rightarrow\omega$ PT in  Zr \cite{Pandey-Levitas-ActaMat-20}.  Based on data, $A\simeq 23$ for $p=p_e$, which we will used due to lack of data for olivine$\rightarrow$spinel PT. In contrast to pressure/stress-induced PT, time is not a parameter in Eq.(\ref{l-g-4}), plastic strain plays a role of a time-like parameter.
Thus, the rate of strain-induced PT  is determined by the rate of plastic deformation.
To reach $c=0.99$, plastic strain $\vep=4.6/A=0.2$, which at strain rate $10\,  s^{-1}$ (or $10^{-4}\,  s^{-1}$) takes just $0.02 s$ (or $20 s$), instead of millions years without plastic strain.
Thus, plastic strain can increase the transformation rate by more than ten orders of magnitude.

\begin{figure}[ht]
\centering
\includegraphics[width= \linewidth]{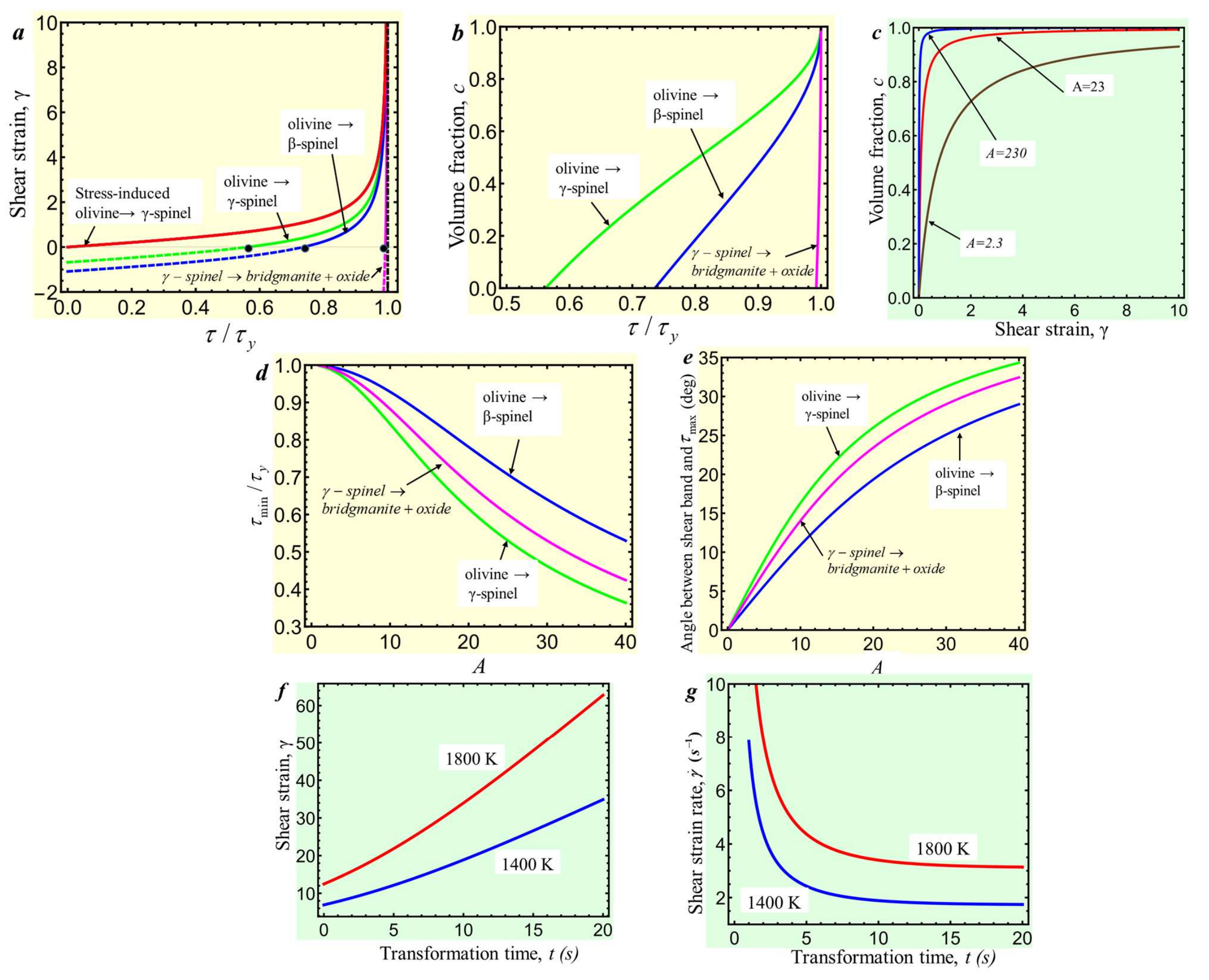}
\caption{{\bf Kinetics of coupled strain-induced phase transformations and TRIP.}
(a) and (b) Shear strain and volume fraction of the high-pressure phase vs. $\tau/ \tau_y $, respectively.
Dots denote shear stress $\tau_{min}$ for initiation of strain-induced PT. Line for the stress-induced PT corresponds to Eq.(\ref{meyers1-207}).
(c) Kinetics of olivine$\rightarrow\gamma$-spinel PT for different kinetic parameters $A$. (d) and (e) Shear stress $\tau_{min}/ \tau_y$ for initiation of strain-induced PT and angle $\alpha$ between the shear-transformation band and direction with maximum shear stress $\tau_{max}$, respectively, vs. kinetic parameter $A$.
Results for  chemical reaction $\gamma$-spinel$\rightarrow$ bridgmanite+ oxide (magnesiowüstite)
are included in (a)-(e) with $A=2.3$ and $\vep_0=0.08$ \cite{Green-07}.
(f) and (g) Shear strain and shear strain rate, respectively,  required to reach temperatures of $1400\,K$ and $1800\,K$ during transformation time $t$ for parameters for the Punchbowl Fault.}
\label{Fig-TRIP}
\end{figure}

Next, we need to find a mechanism for a drastic increase in strain rate and temperature. We suggest that TRIP
caused by olivine$\rightarrow$spinel PT can cause this.
TRIP occurs due to internal stresses caused by volume change during the PT combined with external stresses.
We found (Supplementary Information) an analytical
3D solution, in which  
 the plastic shear $ \gamma $, which is TRIP, is related to
the applied shear stress $ \tau $, the yield strength in shear $ \tau_y $ during PT, and
volumetric transformation strain $ \vep_o $ (see Fig. \ref{Fig-TRIP}(a))  as
\bey
d\gamma/d c =  \frac{2}{\sqrt{3}} \mid  \vep_o  \mid
(\tau /\tau_y)/ \sqrt{1 -  (\tau/ \tau_y)^2}
 \; \rightarrow \;
\gamma \; = \; \frac{2}{\sqrt{3}} c \mid  \vep_o \, \mid
(\tau /\tau_y)/ \sqrt{1 -  (\tau/ \tau_y)^2}. 
\label{meyers1-207}
\eey
Effective transformation volumetric strain $c \vep_o$ during  growth 
of $c$ 
forces plastic strain to restore displacement continuity across an interface (see Fig. \ref{Fig-shear-band}(b) and (c)), and plastic flow takes place at arbitrary (even infinitesimal)
shear stress.  The yield strength in shear $ \tau_y $ during PT is unknown. Atomistic simulations for many materials (e.g., in \cite{Gaoetal-17,Zarkevichetal-18})
show that lattice resistance drops to and even below zero after lattice instability. For strain-induced PT, nanosize nuclei also reduce the yield strength \cite{Karato-20}. We assume that $ \tau_y = const = \sg/\sqrt{3}=173\, MPa$.
For $\tau \rightarrow  \tau_{y}$ (e.g.,  in a shear band), plastic shear tends to infinity (Fig. \ref{Fig-TRIP}(a)).
This is the desired singularity we wanted to find above. Note that our 3D solution has  the proportionality factor  $2\sqrt{3}\simeq 3.4$ times larger
than in the previous 2D treatments \cite{Levitas-JMPS-97,levitas-ijss1998,levitas-ijp2000-2p,levitas+nesterenko+meyers-1998}, which changes the current results qualitatively.
\\
\\
{\bf Self-blown-up deformation-transformation-heating process}
\\
Since PT causes TRIP, which (like traditional plasticity) promotes strain-induced PT, it, in turn, promotes TRIP, and so on, there is positive thermomechanochemical feedback, which we called a self-blown-up deformation-transformation-heating process.
In such a case, Eq.(\ref{l-g-4}) cannot be integrated alone but should be considered together with Eq.(\ref{meyers1-207}).
For shear-dominated flow $\vep= \gamma/\sqrt{3}$, and we obtain (Fig. \ref{Fig-TRIP}(a)-(d))
\bey
\gamma  = 2  \frac{ \mid  \vep_o \mid}{\sqrt{3}}
\frac{\tau}{\tau_y}/ \sqrt{1 - \left (\frac{\tau}{ \tau_y}\right)^2} -\sqrt{3}/A ;
\quad
c= 1-  \frac{3}{2} \sqrt{1 -  \left(\frac{\tau}{ \tau_y}\right)^2}/(\frac{\tau}{ \tau_y}A\mid  \vep_o \mid)= \left(1+ \sqrt{3}/ (A \gamma) \right)^{-1};
\label{l-g-6}
\eey
\bey
 \tau/ \tau_y \geq 1/\sqrt{1+ 4A^2\mid  \vep_o \, \mid^2/9}.
\label{l-g-7}
\eey
Eq.(\ref{l-g-7}) is the criterion for a self-blown-up deformation-transformation-heating process, shown in Fig. \ref{Fig-TRIP}(d) vs. $A$. It is obtained from Eq.(\ref{l-g-6})  and condition $c\geq 0$ or $\gamma \geq 0$. The last expression for $c(\gamma)$ in Eq.(\ref{l-g-6}) is obtained by excluding $ \tau/ \tau_y $ from two previous Eqs.(\ref{l-g-6}).
For olivine$\rightarrow\, \gamma$-spinel PT $ \vep_o=-0.096$  and for olivine$\rightarrow\, \beta$-spinel PT $ \vep_o=-0.06$\cite{Navrotsky-73,Green-07};
 this results in
  $\tau/ \tau_y \geq 0.562$ for $\gamma$-spinel and $\tau/ \tau_y \geq 0.736$ for $\beta$-spinel, which are not very restrictive. Thus, since $\tau/\tau_y=\cos 2 \alpha$, where $\alpha$ is the angle between maximum shear stress and shear band,
the above criterion is met at $\alpha \leq 27.9^o$ for  $\gamma$-spinel  and $\alpha \leq 21.3^o$ for $\beta$-spinel (Fig. \ref{Fig-TRIP}(e)).
We will focus on olivine$\rightarrow\, \gamma$-spinel PT since it has larger TRIP and less restrictive constraints.

To have $\gamma=10$, $\tau/ \tau_y = 0.999939$ and $c=0.9925$; for $\gamma=100$,  $\tau/ \tau_y  =0.999999$ and $c=0.999248$. Thus, for the self-blown-up deformation-transformation process to produce shear $\gamma >10$, one needs $\tau/ \tau_y  =1$, i.e., perfect alignment of maximum shear stress and shear band. This contributes to understanding why the self-blown-up deformation-transformation-heating process and deep-focus earthquakes are relatively rear processes.
Eq. (\ref{l-g-7}) explains extremely large shear strains (sliding) in a fault or friction surface.
Also, since deviatoric  shear strain
is  much larger than $ \vep_o$,
this resolves a puzzle of the deviatoric character of the deep-earthquake source  \cite{Frohlich-89,Zhan-20}.
Note that for very large TRIP shear the term $ -\sqrt{3}/A $ in Eq.(\ref{l-g-6})$_2$ is negligible (Fig. \ref{Fig-TRIP}(a)), i.e., TRIP shear is independent of  any kinetic properties  of strain-induced PT. Also, for $\tau/ \tau_y  \rightarrow 1$,  Eq.(\ref{l-g-6})$_2$ gives $c \rightarrow 1$.
TRIP-induced temperature rise is determined by the  equation
\bey
\rho \nu   \dot{T} h =-4 k (T-T_0)/h+ \tau_y \dot{\gamma} h
,
\label{l-g-8}
\eey
in which for $\tau \rightarrow  \tau_{y}$ we even neglected the  transformation heat to have a conservative estimate.
The solution is
\bey
T=T_0+ (T_s^{tr}-T_0) \left[1-\exp \left(-\frac{4k}{\rho \nu h^2}t \right)\right]; \quad T_s^{tr}=T_0+\frac{ \tau_y \dot{\gamma} h^2}{4k},
\label{l-g-9}
\eey
where $T_s^{tr}$ is the stationary temperature due to TRIP heating. The shear rate to reach temperature $T$ during the PT time $t$, as well as corresponding hear strain $\gamma$ are determined from Eq.(\ref{l-g-9})
\bey
\dot{\gamma}=(T-T_0)\frac{4k}{\tau_y  h^2}\left[1-\exp \left(-\frac{4k}{\rho \nu h^2}t \right)\right]^{-1}; \qquad \gamma=\dot{\gamma} t; \qquad \gamma(t=0)=\frac{\rho \nu}{\tau_y}(T-T_0).
\label{l-g-10}
\eey
Figs. \ref{Fig-TRIP}(f) and (g) exhibit $\dot{\gamma}$ and $\gamma$ required to reach temperatures $1800\,K$ and $1400\,K$ vs. transformation time $t$ for parameters for the Punchbowl Fault. The faster PT is, the smaller shear but larger strain rates are required.   Minimum shears are at $t=0$ (instantaneous PT), $\gamma (1800)=12.5$ and $\gamma (1400)=6.9$  but results in infinite strain rate.  For $t<10\, s$, the desired temperature is reached during transitional heating. For $t>10\, s$,  it is reached by approaching a stationary temperature; that is why the required strain rates approach stationary values.  Based on kinetic estimates in \cite{Karato-20}, time for complete pressure-induced PT at $17\, GPa$  and $1420\,K$ is $10\, s$; strain-induced PT may occur by orders of magnitude faster even at a much lower temperature. Practically,
limitation comes from the required shear (rather than shear rate). Based on Eq.(\ref{l-g-6}), strain $\gamma\geq 10$ requires
$\tau/ \tau_y \geq 0.999939$, i.e., practically perfect alignment of the shear band along the maximum shear direction. Shear rate is calculated by dividing shear by PT time. For $t>1\, s$, shear rate is smaller than $10\, s^{-1}$, and after completing PT it further increases during traditional plastic flow due to $T>T_s$ (Fig. \ref{fig:temp}(b)). For $0.001<t<1\,s$, the shear rate is in the range of
$10-10^4\, s^{-1}$, on the same order of magnitude as it is expected at   $1800\,K$ during traditional plastic flow.

Note that since during PT $\tau/ \tau_y \simeq 1$, traditional plastic flow (which is neglected) should add to TRIP and further increase
both strain rate and temperature.
Thus, TRIP and the self-blown-up deformation-transformation-heating process should lead to temperatures higher than $T_s$ in Fig. \ref{fig:temp}, after which further drastic temperature increase does not need PT and can occur due to traditional plastic flow.
Theoretically, thermoplastic unstable temperature increase above $T_s$ can lead to melting, which is one of the mechanisms of high-strain rate shear localization and deep earthquake  \cite{Frohlich-89,Kanamori-etal-98}. However, due to a strong heterogeneity of earth materials along the  shear band, including nontransforming minerals, melting may not be reached and is not necessary. As it is estimated above, reaching $1800\,K$ is sufficient for achieving strain rates     $10^{2}-10^{4}\, s^{-1}$. We also want to stress that the melting-based mechanism of the deep earthquake is possible in nature only if some other processes (like self-blown-up deformation-transformation-heating)
will increase temperature above $T_s$ (see Supplementary Information).

For laboratory experiments in \cite{green-etal-15}, due to very small thickness of the transformation-deformation band,  temperature
increase due to TRIP is negligible even for $\dot{\gamma}=14\, s^{-1}$ (Supplementary Information).
That means that after completing PT, strain rate should reduce to the value corresponding to nanograined spinel under the same stress and temperature, which is $\sim 10^{-2}\, s^{-1}$, and its contribution to the total shear $\gamma=43$
is small. Thus, this strain is due to TRIP only. Since in the Punchbowl Fault $\gamma=10^6$, the major part of this strain is due to thermoplastic flow after PT and above $T_s$.

 Similar processes are expected in multiple transformation-shear and shear bands (Fig. \ref{fig:cascade}) that find ways through weak obstacles  and may percolate or just increase the total shear-band volume and amplify generated seismic waves.

 In reality, the shear band is not infinite but has a very large (10 to 1000 and larger) ratio of length, at least in the shear direction, to the width.
That is why the above theory is applicable away from the boundary of a band.
When finite-size single or coalesced deformation or transformation-deformation  bands propagate, stresses at their ends are equivalent to those at a dislocation pileup or superdislocation  but at a larger scale \cite{Levitas-etal-PRL-18} and with the total Burgers vector $\gamma h$, which may be huge.
These stresses cause both fast PT and plasticity and further propagation of shear band and trigger initiation of new bands, mostly mutually parallel. Such a stress concentrator is by a factor of $\gamma/\vep_0$, i.e.,  orders of magnitude, stronger than that at the tip of the anticrack  \cite{Burnley-Green-89,Riggs-Green-pileup-05,green-etal-15,green-burnley-89,Kirbyetal-96,green-etal-Science-13,Green-07,Green-17} and much more effective in spreading  transformation-deformation bands at the higher, microscale. The resulting propagating thermoplastic band can pass through non-transforming minerals and extend  outside of metastable olivine wedge. Indeed, it was demonstrated in \cite{Green-17} that the fault originated in metastable  Mg$_2$GeO$_4$ olivine during its transformation to spinel propagated through previously transformed spinel.

To summarize, our  quantitative mechanisms of very fast localized thermoplastic flow and PT consist of several interrelated steps shown in Fig. \ref{fig:cascade} and contain several conceptually important points:
(a) Proof that plastic flow alone cannot lead to localized in mm-scale band
heating, that is why PT is required; (b) Substitution of stress-induced PT with plastic strain-induced PT, which was not previously used in geophysics and leads to completely different kinetic description;
(c) Transition to dislocation flow with strong stress concentrators is required to substitute stress-induced PT with barrierless and fast plastic strain-induced PT; (d)     Strain-induced PT in a shear band generates severe (singular) TRIP
and heating, which in turn produces  strain-induced PT and so on, resulting in the self-blown-up PT-TRIP-heating process    due to positive thermomechanochemical feedback.
(e) This leads to the heating above the  temperature, exceeding unstable stationary temperature $T_s=1400-1800\,K$, after which
further heating in a shear band occurs due to traditional thermoplastic flow.
Achieving  $T=1800\, K$ is sufficient to reach $\dot{\vep}(T)=10-10^{3}\, s^{-1}$ and   generate seismic waves.
(f)  These processes repeat themselves at larger scale.

Lack of any of these processes due to not meeting the required conditions (e.g., proper orientation or path with a small content of stronger phases) may lead to inability to   reach very fast localized PT and plastic flow and cause an earthquake, which explains that earthquakes are relatively rare events.
Similarly,  lack of seismic activity below 660 km, where endothermic
 and slow disproportionation reaction from $\gamma$-spinel   to bridgmanite)+ oxide (magnesiowüstite) occurs, can be explained (see Supplementary Information).

 Relatively small shear strain in laboratory experiment  \cite{Riggs-Green-pileup-05} ($\gamma=43$ vs. $\gamma=10^6$ in nature)
is because  the temperature cannot grow due to an extremely thin  band, processes in the third column in Fig. \ref{fig:cascade} are absent, and TRIP occurs only (see Supplementary Information).
A very rare occurrence of such bands in a laboratory  is related to   small sample size and low probability of realization of sequences of all the processes  in  Fig. \ref{fig:cascade}, as well as to small temperature window when stress-induced anticrack cannot appear but strain-induced bands can.
 Our  Eq.(\ref{Eq-1})  relates the change in strain rate with respect to the initial one before localization. That is why the final strain rate is distributed with depth similar to the initial strain rate before localization. This is consistent with the correlation between seismicity in transition zone and strain rate before localization \cite{Billen-20}.
 Eq. (\ref{l-g-7}) explains extremely large shear strains (sliding) in a fault or friction surface.
Since   shear strain
is  much larger than the volumetric strain,
this resolves a puzzle of the deviatoric character of the deep-earthquake source  \cite{Frohlich-89,Zhan-20}.

Our findings change the main concepts in studying the initiation of the deep-focus earthquakes and PTs during plastic flow in geophysics in general.  They will be elaborated in much more detail using modern computational multiscale
approaches for studying coupled PTs and plasticity \cite{Levitas-MT-19}.  Introducing strain-induced PT and the self-blown-up transformation-TRIP-heating process
 may change the interpretation of various geological phenomena.
 In particular, they may explain
possibility of the appearance of microdiamond directly in the cold Earth crust within shear bands \cite{Gaoetal-17} during tectonic activities without subduction to the high-pressure and high-temperature mantle and uplifting. Developed theory of the self-blown-up transformation-TRIP-heating process is applicable outside geophysics for various processes in materials under pressure and shear, e.g., for new routes of material synthesis, friction and wear under high load, penetration of the projectiles and meteorites, surface treatment, and severe plastic deformation and mechanochemical technologies   \cite{levitas-mechchem-04,levitas-prb-04,Edalati-Horita-16,Levitas-JPCM-18,Levitas-MT-19,koch-1993,zharov-1994,Takacs-2013,Balazetal-2013}.
\\\\
{\bf Acknowledgements.}
Support from
NSF (CMMI-1943710 and DMR-1904830),    and Iowa State University (Vance Coffman
Faculty Chair Professorship) is greatly appreciated.

\newpage
\renewcommand{\thepage}{\arabic{page}}

\baselineskip14pt
\belowdisplayskip10pt
\belowdisplayshortskip10pt
{

}

\newpage

\bec
{\bf \large  Supplementary Information}
\\
{\bf \large  Resolving puzzles of the phase-transformation-based mechanism of the deep-focus earthquake }
\\
Valery I. Levitas\\
{\it
Iowa State University, Departments of Aerospace Engineering and
Mechanical Engineering,  Ames, Iowa 50011, USA\\
Ames Laboratory,
Division of Materials Science and Engineering,  Ames, IA, USA}

\eec

\section{Kinetics of plastic strain-induced phase transformations}
The strain-controlled kinetic equation derived in \cite{levitas-prb-04,levitas-mechchem-04} using the main conceptual
results from the nanoscale modeling of nucleation at the dislocation pileup and micromechanical treatment is
\bey
\frac{d c}{d q} =
a \; {\left(1-c \right)^s }\,\frac{\sg_{y2}^w}{\sg_a}\,
\frac{p - p_\vep^d}{p_h^d - p_\vep^d}H(p - p_\vep^d)
 -
b \; c^s \, \frac{\sg_{y1}^w}{\sg_a}\,
\frac{p_\vep^r - p}{p_\vep^r - p_h^r} H(p_\vep^r - p); \quad \sg_a = c   \sg_{y1}^w   +
\left(1   -   c \right) \sg_{y2}^w.
\label{l-g-4-a}
\eey
Here, $p$ is the pressure, $c$ is the volume fraction of a high-pressure phase,
$ \sg_{yi} $ is the yield strength of $ i$-th phase;
$p_{\vep}^d$ and $p_{\vep}^r$  are the minimum pressure at which the direct strain-induced phase transformation (PT) may occur and maximum pressure at which the reverse strain-induced PT proceeds, respectively, $H$ is  the Heaviside step function used to impose criteria for the direct ($p  >  p_\vep^d$) and reverse ($p<p_\vep^r$) strain-induced PTs;
$p_h^d$ and $p_h^r$ are the pressures for the direct and reverse PTs under hydrostatic loading;   symbols $a$, $b$, $s$, and $w$ are material parameters.   Eq.(\ref{l-g-4-a}) includes
 the possibility of direct and reverse PTs and
 the different plastic strain in each phase due to different $ \sg_{yi} $.

 We do not consider strain-induced reverse spinel$\rightarrow$olivine PT because resultant nanograin spinel deforms dominantly by grain-boundary sliding, which does not produce stress concentrators inside the grains.
 The difference in yield strength of phases is neglected for compactness and  $s=1$.
Then Eq.(\ref{l-g-4-a}) reduces to
\bey
\frac{d c}{d \vep} = A {\left(1-c \right)}  \qquad  {\rm for} \; p>p_\vep^d (T); \quad A:=
a \frac{p - p_\vep^d (T)}{p_h^d (T) - p_\vep^d (T)} \quad \rightarrow \quad   c=1- \exp(-A \vep),
\label{l-g-4-sup}
\eey
which is Eq. (\ref{l-g-4}) in the main text.

\section{Analytical 3D solution for transformation induced plastic shear in a transformation-shear band}

\begin{figure}[ht]
\centering
\includegraphics[width=0.7 \linewidth]{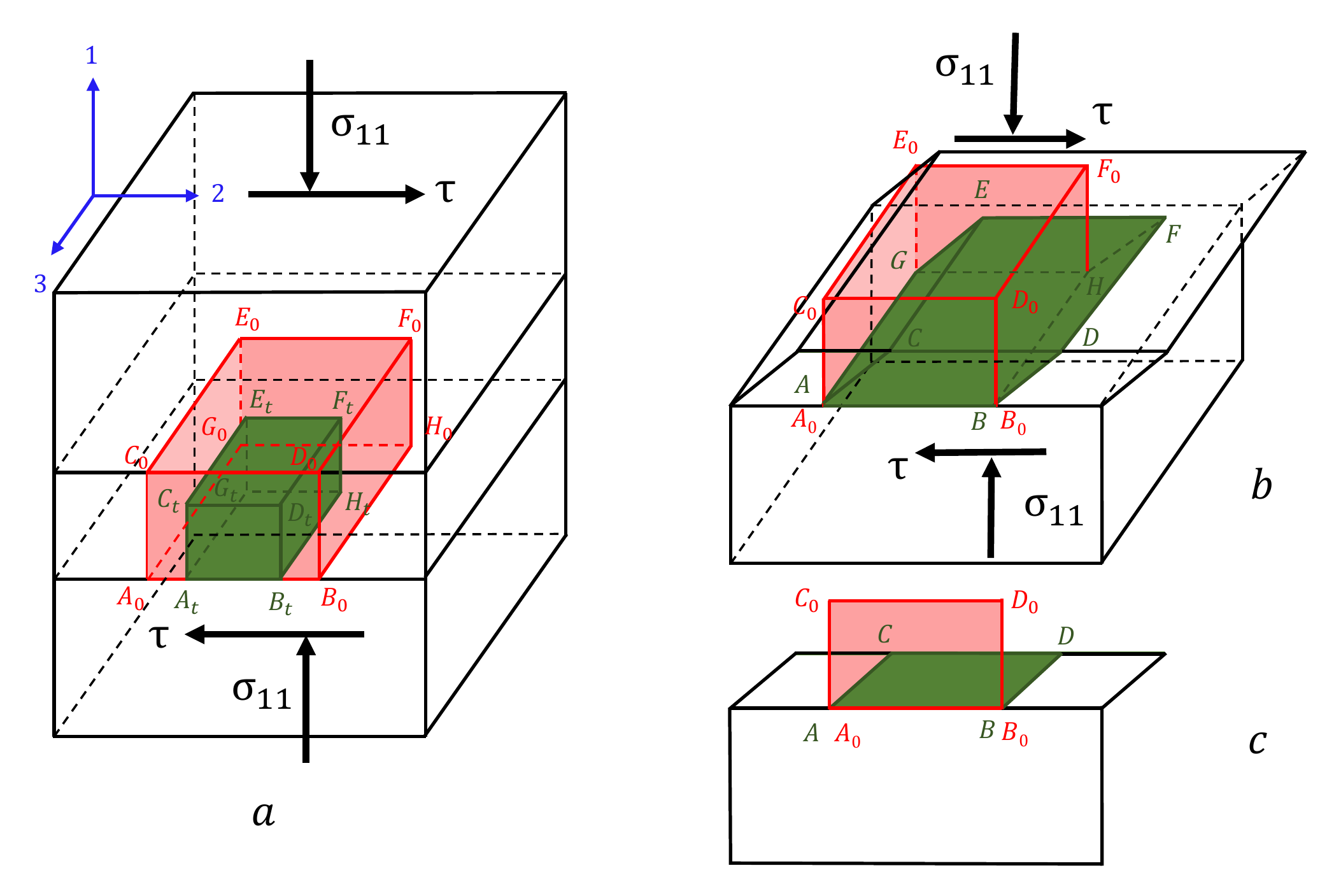}
\caption{{\bf Schematics of processes in a  transformation-deformation-heating band within a rigid  space.} (a) Part of a band before PT (red) and after PT and isotropic transformation strain (green) is shown. (b)
To satisfy the continuity of displacements across the shear-band boundary and rigid space outside the band, additional TRIP
 develops, leading  to deformation of the green rectangular $A_tB_tG_tH_t$ to $ABGH$ that coincides with $A_0B_0G_0H_0$ and to   large plastic shear. (c) 2D view (along axis 3) of (b).}
\label{Fig-shear-band-supp}
\end{figure}

To model the transformation-deformation band in olivine, we consider an infinite space within which localized plastic deformation and PT occur (Fig. \ref{Fig-shear-band-supp}).
TRIP occurs due to internal stresses caused by volume change during the PT combined with external stresses.
We found simple analytical
solutions for PT in a plastic shear band at small  \cite{Levitas-JMPS-97,levitas-ijp2000-2p}  and large \cite{levitas+nesterenko+meyers-1998}  strains in the 2D formulation. Here, we find the first 3D analytical solution.
We consider the homogeneous state of the space and band before strain localization and phase transformations as the reference state. Then change in elastic strains is small and can be neglected. Thermal strains are also minor compared to plastic and transformational strains and will be neglected as well. Deformations outside the band are negligible, i.e., rigid space is considered. The stress-strain state within the band is homogeneous.
For compactness and transparency, we will use small-strain formalism, while the final results will be valid for large plastic and small transformational strains, which is the case for olivine$\rightarrow\, \gamma$-spinel PT (volumetric transformation strain for complete PT $ \vep_o=-0.096$)  and for olivine$\rightarrow\, \beta$-spinel PT ($ \vep_o=-0.06$)\cite{Navrotsky-73,Green-07}.

Let us choose the orthogonal coordinate system with axis 1 directed along
the normal $\, {\fg n} \,$ to the shear band and axes 2 and 3 parallel to the shear band.
We divide six components of any symmetric tensor, e.g., stress $\fsg $, into two parts:
\bey
\fsg_n \, := \,
\left( \begin{array}{ccc}
       \sg_{11} & \sg_{12} & \sg_{13} \\
       \sg_{21} & 0        & 0        \\
       \sg_{31} & 0        & 0
       \end{array} \right) \; = \left( \begin{array}{ccc}
       \sg_{11} & \tau & 0 \\
       \tau & 0        & 0        \\
       0 & 0        & 0
       \end{array} \right) ;
\qquad \qquad
\fsg_{||} \, := \,
\left( \begin{array}{ccc}
       0        & 0        & 0        \\
       0        & \sg_{22} & \sg_{23} \\
       0        & \sg_{32} & \sg_{33}
       \end{array} \right) .
\label{ijss1-159}
\eey
Components  $\fsg_n$ are stresses acting at the surface of the shear band, i.e., components of the traction vector; components $\fsg_{||}$ are in-band stresses. Also, we chose axis 2 along the applied shear stress $\sg_{12}=\tau$, i.e.,  $\sg_{13}=0$.
Components $\fsg_n$ within the band are equal to corresponding components in the space due to traction continuity conditions. Then they are equal to applied stresses and are considered independent of time during phase transformation.

The total strain $\fvep^{tot}$ consists of plastic  $\fvep$ and transformational $\fvep^{t}$
parts:
\bey
\fvep^{tot} \, := \,
\left( \begin{array}{ccc}
       \vep_{11}^{tot} & \vep_{12}^{tot} & \vep_{13}^{tot} \\
       \vep_{21}^{tot} & \vep_{22}^{tot}        & \vep_{23}^{tot}        \\
       \vep_{31}^{tot} & \vep_{32}^{tot}       & \vep_{33}^{tot}
       \end{array} \right) \; = \left( \begin{array}{ccc}
       \vep_{11}  & \vep_{12}  & \vep_{13}  \\
       \vep_{21}  & \vep_{22}         & \vep_{23}        \\
       \vep_{31}  & \vep_{32}        & \vep_{33}
       \end{array} \right) \;+
       \left( \begin{array}{ccc}
       \vep_t & 0 & 0 \\
       0 &  \vep_t         & 0        \\
      0 & 0        &  \vep_t
       \end{array} \right) ,
\label{earth-1s}
\eey
where transformation strain is a spherical (pure volumetric) tensor.
Decomposing Eq.(\ref{earth-1s})  in normal and in-band parts, we obtain
\bey
\fvep^{tot}_n \, := \,
\left( \begin{array}{ccc}
       \vep_{11}^{tot} & \vep_{12}^{tot} & \vep_{13}^{tot} \\
       \vep_{21}^{tot} & 0        & 0        \\
       \vep_{31}^{tot} & 0      & 0
       \end{array} \right) \; = \left( \begin{array}{ccc}
       \vep_{11}  & \vep_{12}  & \vep_{13}  \\
       \vep_{21}  & 0         & 0        \\
       \vep_{31}  & 0        & 0
       \end{array} \right) \;+
       \left( \begin{array}{ccc}
       \vep_t & 0 & 0 \\
       0 &  0        & 0        \\
      0 & 0        & 0
       \end{array} \right) ;
\label{earth-2s}
\eey
\bey
\fvep^{tot}_{||} \, := \,
\left( \begin{array}{ccc}
      0 & 0 & 0\\
     0 & \vep_{22}^{tot}        & \vep_{23}^{tot}        \\
     0 & \vep_{32}^{tot}       & \vep_{33}^{tot}
       \end{array} \right) \; = \left( \begin{array}{ccc}
       0  & 0  & 0  \\
      0  & \vep_{22}         & \vep_{23}        \\
      0  & \vep_{32}        & \vep_{33}
       \end{array} \right) \;+
       \left( \begin{array}{ccc}
     0 & 0 & 0 \\
       0 &  \vep_t         & 0        \\
      0 & 0        &  \vep_t
       \end{array} \right) .
\label{earth-3s}
\eey
Due to continuity of displacements across the shear-band boundary and rigid space outside the band (i.e., the coherent boundary between shear-band and the rest of the space), we obtain
  \bey
\fvep^{tot}_{||} \, := \, \fg 0  \quad \rightarrow \quad
\fvep_{||} \, := \,
\left( \begin{array}{ccc}
      0 & 0 & 0\\
     0 &- \vep_t          & 0       \\
     0 &0       & - \vep_t
       \end{array} \right) .
\label{earth-4s}
\eey
Eq.(\ref{earth-4s}), which
was derived in \cite{levitas-ijss1998,levitas-ozsoy-IJP-09},  directly follows from the Hadamard compatibility condition \cite{Truesdell-Toupin} across a coherent boundary. Geometrically, it means that the boundary is undeformed and, due to homogeneity of the strain state within a band, in-band strains are absent.

It follows from Eq.(\ref{earth-4s}) and the plastic incompressibility
\bey
\vep_{11}+\vep_{22}+\vep_{33}=0 \quad \rightarrow  \quad \vep_{11}= 2 \vep_t .
\label{earth-5s}
\eey
The von Mises yield condition
\bey
|\fg S|:= \sqrt{S_{11}^2 + S_{22}^2+S_{33}^2+ 2 \sg_{12}^2+ 2 \sg_{13}^2+ 2 \sg_{23}^2}= \sqrt{\frac{2}{3}} \sg_y= \sqrt{2} \tau_y,
\label{earth-6s}
\eey
where $\sg_y= \sqrt{3} \tau_y$  and $\tau_y$ are the yield strengths in compression and shear, respectively,  $\fg S= \fsg - \frac{1}{3} p \fg I$ is the deviatoric stress, and $\fg I $ is the unit tensor.
The yield strength during PT is unknown, and based on the discussion in the main text, we will consider it a constant.
Associated with   the von Mises yield condition plastic flow rule is the proportionality between plastic strain rate and deviatoric stress tensors:
\bey
\left( \begin{array}{ccc}
       2\dot{\vep}_t  & \dot{\vep}_{12}  & \dot{\vep}_{13}  \\
       \dot{\vep}_{21}  & -\dot{\vep}_t         & 0        \\
       \dot{\vep}_{31}  & 0        & -\dot{\vep}_t
       \end{array} \right) \;=\, h \,
       \left( \begin{array}{ccc}
       S_{11} & \tau & 0 \\
       \tau  &  S_{22}         & \sg_{23}        \\
      0 & \sg_{32}        & S_{33}
       \end{array} \right) ,
\label{earth-7s}
\eey
where $h$ is the proportionality factor. It follows from Eq.(\ref{earth-7s})
\bey
 \dot{\vep}_{13}= \dot{\vep}_{31}=0; \qquad  \sg_{23}= \sg_{32}=0; \qquad  S_{22}=S_{33}=- 0.5 S_{11}.
\label{earth-8s}
\eey
Designating $\gamma=2 \vep_{12}$ and utilizing Eq.(\ref{earth-8s}), plastic flow rule simplifies to
\bey
\left( \begin{array}{ccc}
       2\dot{\vep}_t  & 0.5 \dot{\gamma}  & 0  \\
        0.5 \dot{\gamma}  & -\dot{\vep}_t         & 0        \\
       0 & 0        & -\dot{\vep}_t
       \end{array} \right) \;=\, h \,
       \left( \begin{array}{ccc}
       S_{11} & \tau & 0 \\
       \tau  &  -0.5 S_{11}         & 0        \\
      0 & 0        &-0.5 S_{11}
       \end{array} \right) .
\label{earth-9s}
\eey
Eq.(\ref{earth-9s}) contains just two independent equations
\bey
0.5  \dot{\gamma}= h \tau ; \qquad   2\dot{\vep}_t= h S_{11}\quad \rightarrow \quad \frac{ \dot{\gamma}}{\dot{\vep}_t}= 4\frac{ \tau}{S_{11}}.
\label{earth-10s-a}
\eey
To exclude $S_{11}$, we utilize the plasticity condition Eq.(\ref{earth-6s})
\bey
\sqrt{3/2 \, S_{11}^2 + 2 \tau^2}= \sqrt{2} \tau_y \quad \rightarrow \quad S_{11}=sign(\dot{\vep}_t)\frac{2}{\sqrt{3}} \sqrt{\tau_y^2- \tau^2}.
\label{earth-11s-a}
\eey
Substituting Eq.(\ref{earth-11s-a}) in Eq.(\ref{earth-10s-a}), we obtain
\bey
  \dot{\gamma}=2 \sqrt{3} |\dot{\vep}_t | \frac{ \tau/\tau_y}{ \sqrt{1- \left(\tau/\tau_y \right)^2}}.
\label{earth-12s-a}
\eey
Since during PT the volumetric transformation strain is $\vep_o c$, where $c$ is the volume fraction of spinel, then
$\dot{\vep}_t= \frac{1}{3}\vep_o \dot{c}$, and Eq.(\ref{earth-12s-a}) takes its final form
\bey
  \dot{\gamma}=\frac{2 \sqrt{3}}{3} |\vep_o ||\dot{c} | \frac{ \tau/\tau_y}{ \sqrt{1- \left(\tau/\tau_y \right)^2}} \quad \rightarrow
  \quad \frac{d \gamma}{dc}= sign (dc) \frac{2 \sqrt{3}}{3} |\vep_o | \frac{ \tau/\tau_y}{ \sqrt{1- \left(\tau/\tau_y \right)^2}} .
\label{earth-14s-a}
\eey
Eq.(\ref{earth-14s-a}) represents an explicit expression for plastic shear strain rate induced by PT, i.e., TRIP shear. During both direct and reverse PT, TRIP shear increases independent of the sign of the volumetric transformation strain is $\vep_o $.
While Eq.(\ref{earth-14s-a}) has the same form as previous 2D solutions, the proportionality factor is $2\sqrt{3}\simeq 3.4$ times larger
than for 2D treatment.
Eq.(\ref{earth-14s-a}) is used as Eq. (5) in the main text.

For completeness,  pressure $p$ is determined from the equation
\bey
 p = 3 (\sg_{11} - S_{11}).
 \label{earth-15s-a}
\eey

\section{Heat transfer analysis of laboratory experiments \cite{Riggs-Green-pileup-05,green-etal-15}}

\begin{figure}[ht]
\vspace{5mm}
\centering
\includegraphics[width=0.7 \linewidth]{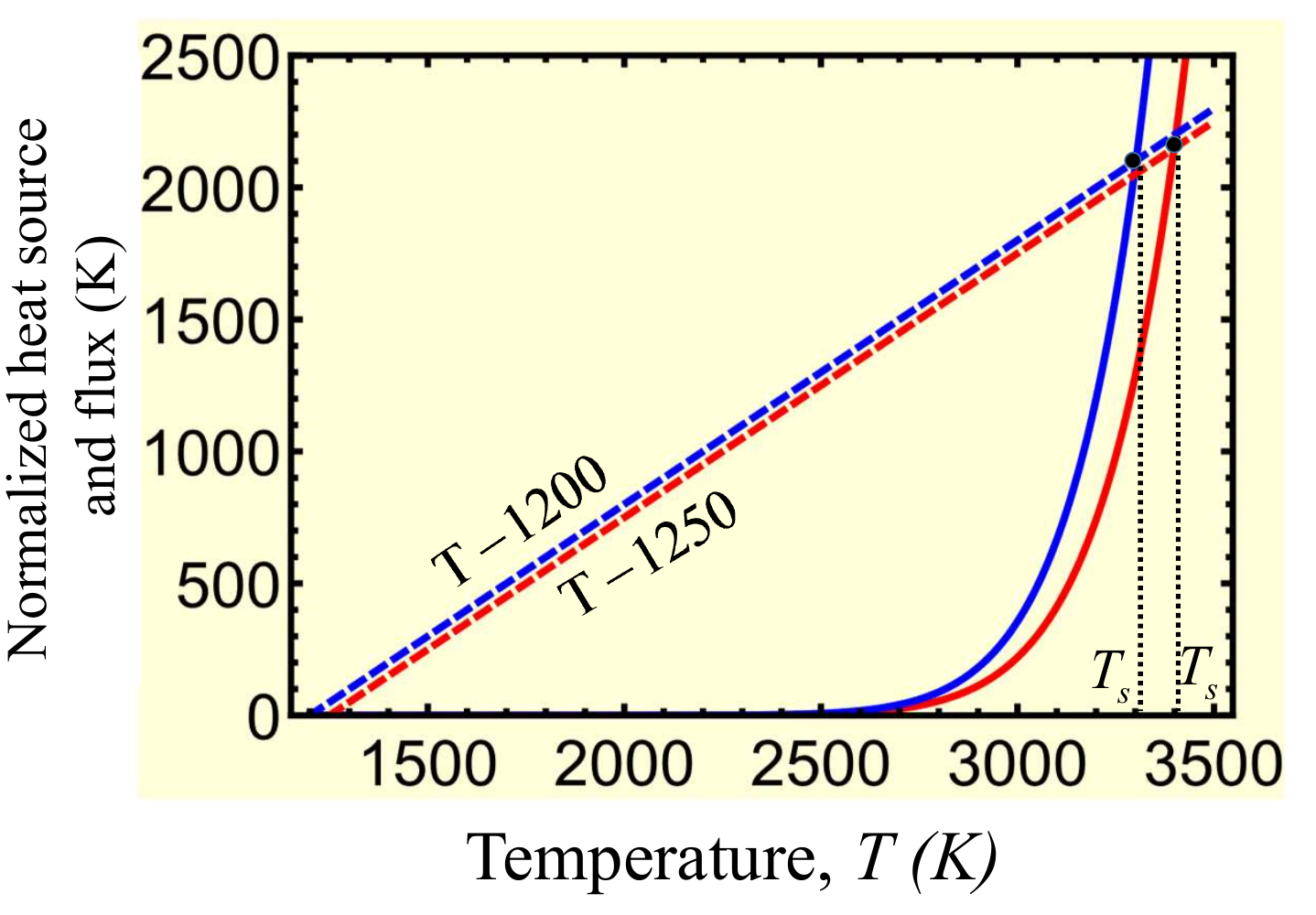}
\caption{{\bf Analysis of experiments  in \cite{Riggs-Green-pileup-05,green-etal-15}.}
Plots of both sides of Eq.(3) in the main text for stationary temperature, namely the straight line  related to the heat flux from the band and the term associated with the plastic dissipation, for two  different sets of experiments in \cite{Riggs-Green-pileup-05,green-etal-15}.
Blue lines correspond to the experiment at $T_0=1200\, K$ and red lines are for $T_0=1250\, K$. Since unstable stationary temperatures $T_s$ for both experiments are very high, they cannot be reached by thermoplastic flow alone, and PT with TRIP is required.}
\label{Fig-exp-Green}
\vspace{5mm}
\end{figure}

Substituting in Eq.(3) of the main text data for Mg$_2$GeO$_4$ from \cite{Riggs-Green-pileup-05}, namely (sample GL707),
$\dot{\vep}_0=2 \times 10^{-4}\, s^{-1}$, $T_0=1250\, K$,  $\sg=1589\, MPa$, and
$h=10^{-7}\, m$,  as well as from
\cite{green-etal-15},
$\dot{\vep}_0=10^{-4}\, s^{-1}$, $T_0=1200\, K$, $\sg=1804\, MPa$, and $h=0.7 \times 10^{-7}\, m$,
we obtain $T_s=3398\,K$ for the first case and $T_s=3302\,K$ for the second case (see Fig. \ref{Fig-exp-Green}). Due to very small shear band thickness in the laboratory experiments, these values are extremely high, far away from the region of stability of spinel, and well above the melting temperature. Since no traces of reverse PT to olivine and melting were observed in   \cite{Riggs-Green-pileup-05,green-etal-15},
these temperatures were not reached, and no thermoplastic shear localization is possible without PT, TRIP, and self-blown-up deformation-transformation-heating process.

However, even with TRIP, substituting in Eq.(9) of the main text  data from the same laboratory experiment \cite{green-etal-15}
$h=0.7 \times 10^{-7}\, m$,  $\dot{\gamma}=14\, s^{-1}$ (see the main text), and maximum $\tau_y=300\, MPa$ from Fig. S2 in \cite{green-etal-15}, we obtain that the maximum (stationary) temperature increase is just $1.3 \times 10^{-6}\, K$. This should not be surprising because thickness $h=70 \, nm$ in a laboratory experiment is  smaller than in Earth $h=4 \, mm$ by a factor of  57143.
Since stationary temperature increment is proportional to $h^2$, for $h=4 \, mm$, $\dot{\gamma}=14\, s^{-1}$, and $\tau_y=300\, MPa$, it would be $4.33 \times 10^{3}\, K$.
Thus, in laboratory experiments on Mg$_2$GeO$_4$ \cite{green-etal-15} temperature increase in the transformation-shear band was absent.

In \cite{green-etal-15}, adiabatic approximation was used to  estimate maximum shear stress and internal friction coefficient from the condition that temperature increment does not exceed $230\, K$, maximum increment to reach the olivine-spinel  phase equilibrium
temperature. A paradoxical result was that the estimated shear stress and friction coefficient were an order of magnitude lower than directly measured. The reason for this paradox is in adiabatic approximation; when heat flux from the shear band is included, the temperature increase is negligible for any reasonable shear resistance and does not restrict the internal friction stress.
This also means that the sliding should drastically reduce after completing PT; that is why shear in the Punchbowl Fault, $\gamma=10^6$, is drastically larger than in  the laboratory, $\gamma=43$. Consequently, processes in the third column in Fig. 2 in the main text
are absent in laboratory experiments and cannot be verified due to small shear band thickness.

Similarly,  drastic heating leading to melting and dissociation is predicted   in \cite{Kanamori-etal-98} using adiabatic approximation.
When heat flux is included, conditions for melting are quite restrictive.

\section{Conditions for unstable heating in the shear band due to thermoplastic flow alone}

\begin{figure}[ht]
\centering
\includegraphics[width=0.7 \linewidth]{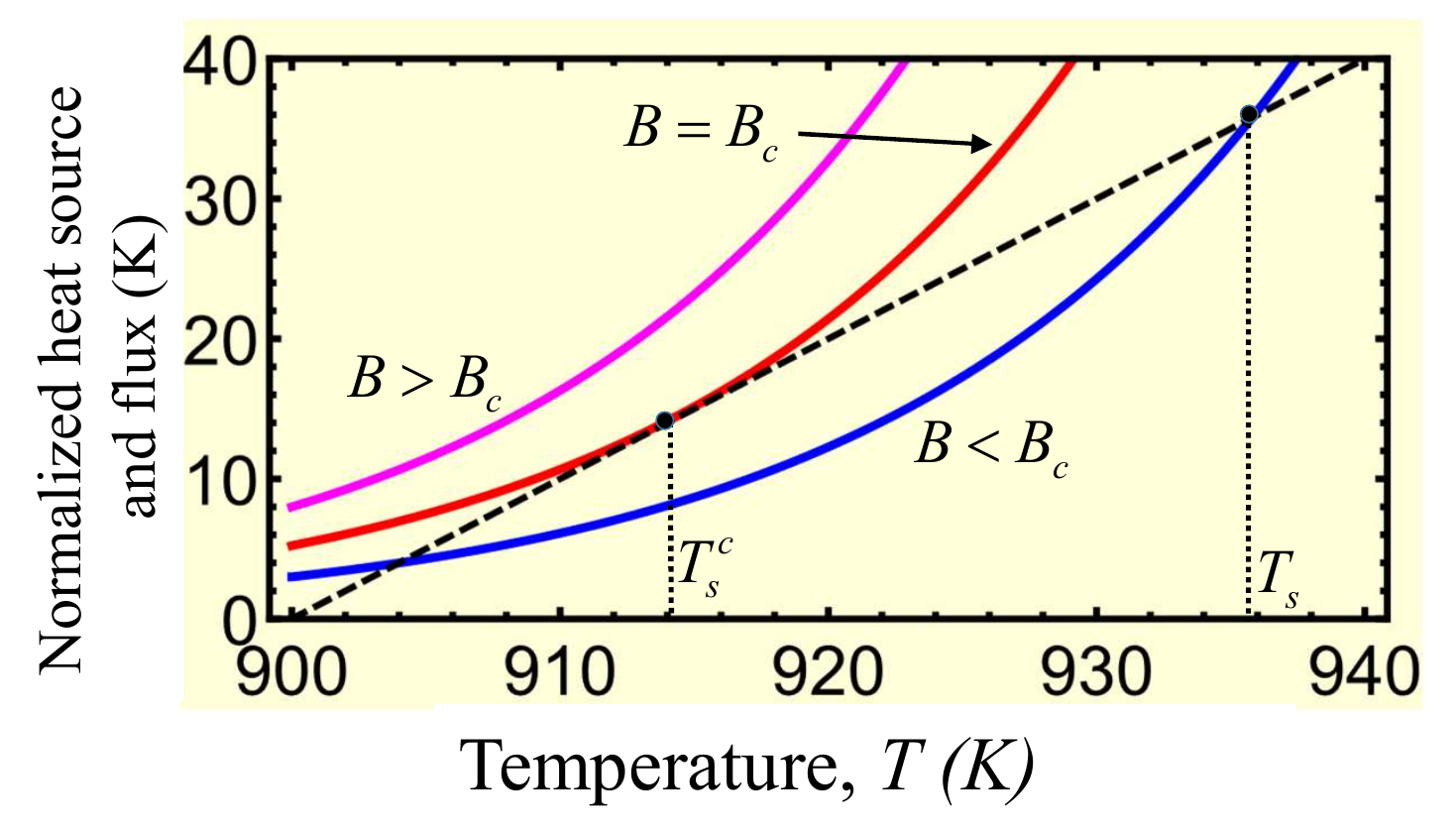}
\caption{{\bf Geometric interpretation of the condition for the unstable heating in the shear band due to thermoplastic flow alone.}
For relatively small parameter $B<B_c$, there are two stationary temperatures, the stable one close to $T_0$ and the unstable one $T_s$.
During thermoplastic heating without PT, solution stacks in the stable stationary temperature. For the critical $B=B_c$, both stationary solutions coincide and are unstable, i.e., unlimited heating should occur. For $B>B_c$,  a static solution does not exist, and unlimited heating should occur.
}
\label{Fig-Ts-schematic}
\end{figure}

Let us rewrite Eq. (3) from the main text for the stationary temperature during plastic flow alone  (i.e., without PT) in the more compact form
\bey
T_s-T_0= B  \exp[-Q_r(T_s^{-1}-T_0^{-1})];        \qquad              B=0.25 h^2  \sg \dot{\vep}(T_0)/ k.
\label{earth-10s}
\eey
As it follows from Fig. 3(b) in the main text and Fig. \ref{Fig-Ts-schematic} here, for small $B$  Eq.(\ref{earth-10s}) has two solutions: one of the solutions with $T \simeq T_0$ is stable and another one with $T_s\gg T_0$ is unstable. This means heating in the shear band for small $B$ is impossible without extra heat sources leading    $T>T_s$. With increasing $B$, the first stable solution slightly exceeds $T_0$ while $T_s$ reduces much faster. At some critical $B=B_c$ and $T=T_c$ both solutions coincide, plastic dissipation exceeds the heat flux from the band for all temperatures (excluding $T=T_c$), the  stationary solution is unstable, and unlimited heating occurs for any infinitesimal perturbation. For critical $B$, derivatives of both sides of Eq.(\ref{earth-10s}) coincide, i.e.,
\vspace{5mm}
\vspace{5mm}

\begin{figure}[ht]
\vspace{5mm}
\centering
\includegraphics[width=0.6 \linewidth]{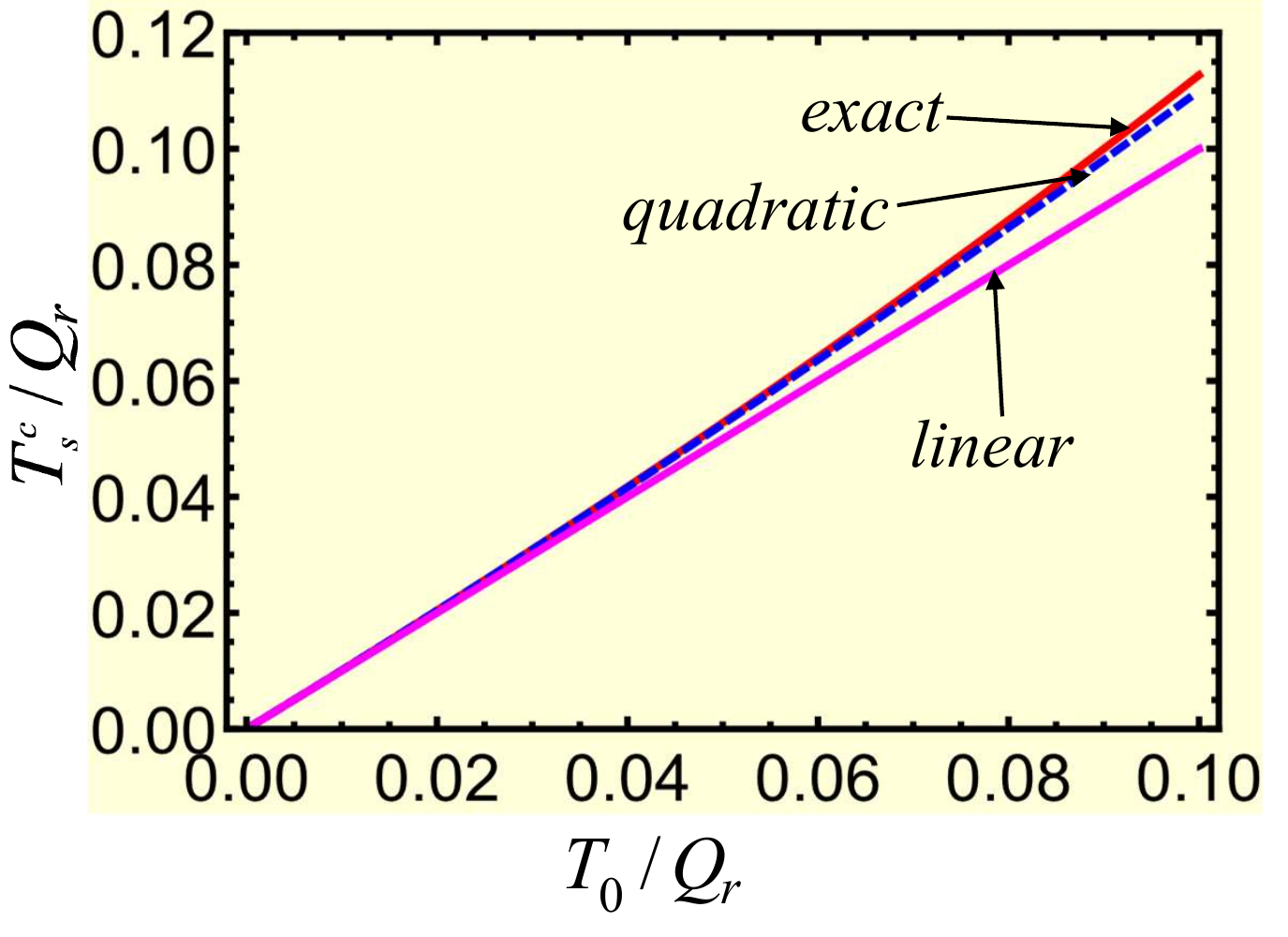}
\caption{{\bf Plots for $T_s^c/Q_r$  vs. $T_0/Q_r$.} Exact dependence, as well as  linear and quadratic approximations, are shown.}
\label{Fig-Tsc}
\vspace{-2mm}
\end{figure}

\begin{figure}[ht]
\vspace{5mm}
\centering
\includegraphics[width=0.6 \linewidth]{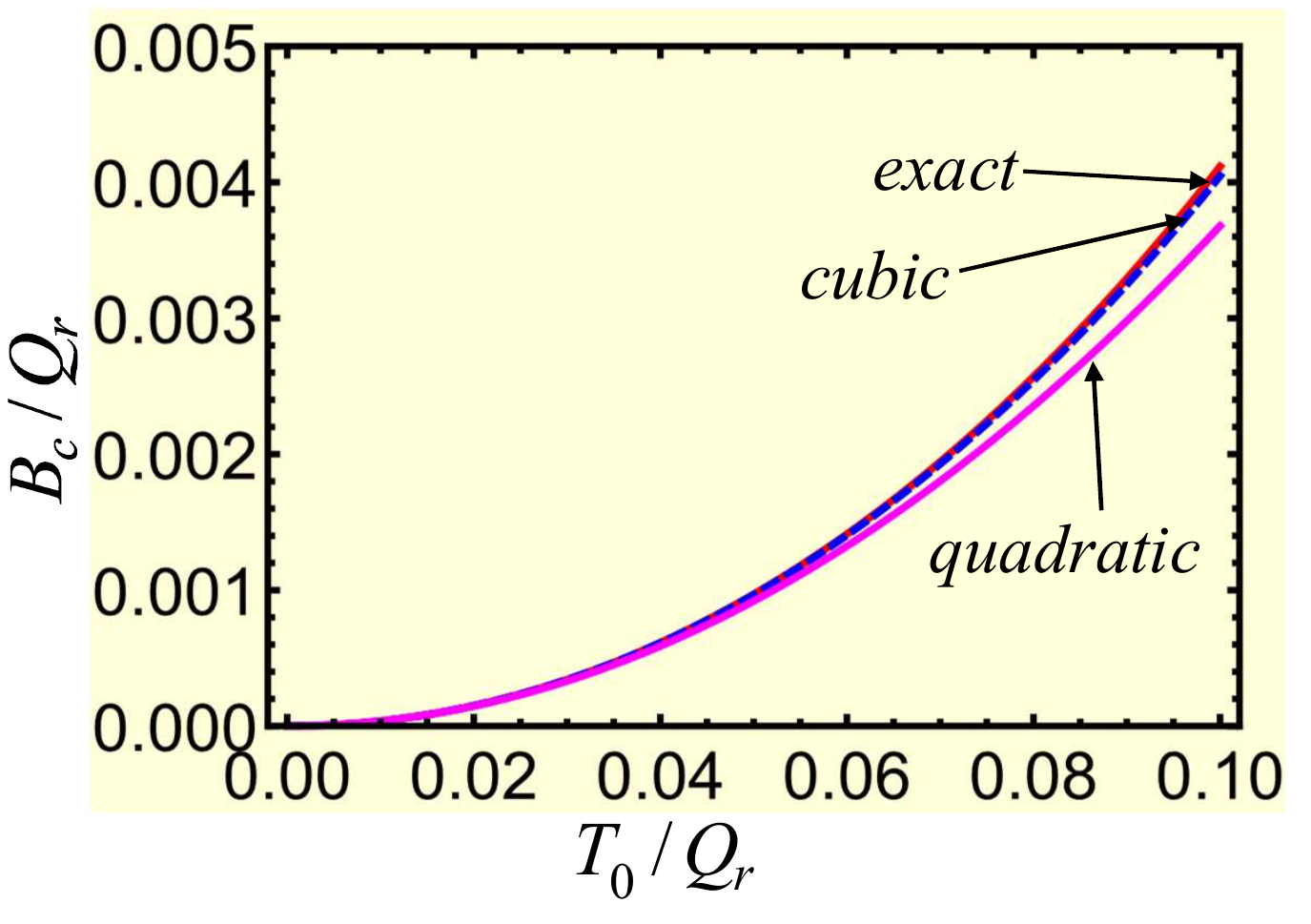}
\caption{{\bf Plot for  $B_c/Q_r$ vs. $T_0/Q_r$.} The exact relationship and  quadratic and cubic approximations are presented.}
\label{Fig-Bc}
\end{figure}

\bey
T_s^2= B Q_r \exp[-Q(T_s^{-1}-T_0^{-1})] .
\label{earth-11s}
\eey
Excluding exponent from Eqs. (\ref{earth-10s}) and (\ref{earth-11s}), we obtain a simple equation
\bey
T_s^2=  Q_r (T_s-T_0) ,
\label{earth-12s}
\eey
with the relevant solution
\bey
T_s^c=  \frac{Q_r }{2} \left( 1-\sqrt{1-4\frac{T_0}{Q_r}}  \right)   \simeq Q_r \left[ \frac{T_0}{Q_r}+ \left(\frac{T_0}{Q_r} \right)^2+2 \left(\frac{T_0}{Q_r} \right)^3  \right]  ,
\label{earth-14s}
\eey
where due to the smallness of  $T_0/Q_r$ the Taylor series is used. Substituting exact $T_s$ from Eq.(\ref{earth-14s}) in Eq. (\ref{earth-10s}), we obtain critical value of $B$:
\bey
B_c= \frac{Q_r }{2} \left( 1-\sqrt{1-4\frac{T_0}{Q_r}} -2\frac{T_0}{Q_r} \right)   \exp\left[-\frac{2}{ 1+\sqrt{1-4\frac{T_0}{Q_r}}}\right]       \simeq
\frac{Q_r}{e} \left[  \left(\frac{T_0}{Q_r} \right)^2+ \left(\frac{T_0}{Q_r} \right)^3  \right],
\label{earth-15s}
\eey
where $e=2.718...$ is the Euler's number. Figs. \ref{Fig-Tsc} and \ref{Fig-Bc}  show plots for $T_s/Q_r$ and $B_c/Q_r$ vs. $T_0/Q_r$. It is clear that for $0\leq T_0/Q_r\leq 0.1$,  $T_s/Q_r$  is well approximated by quadratic function and reasonably good by the linear one. The cubic approximation is not distinguishable from the
exact equation. Similar $B_c/Q_r$  is very good approximated by a cubic polynomial and reasonably good by the quadratic one.
\vspace{5mm}

For data that we used for the Punchbowl Fault, $k= 2.4 \times 10^{-6} MPa \, m^2/(s \, K)$, $Q_r=58,333 K$, $T_0=900\, K$,  $h=4 \times 10^{-3}m$,
  $\sg=300 \, MPa$, $\dot{\vep}(T_0)=10^{-14}$ - $10^{-10}\, s^{-1}$, we have $T_0/Q_r=0.0154$ and $B=5 \times 10^{-12}-5 \times 10^{-8}K$, while $T_s=914.33\, K$ and
$B_c=5.19 \,K$, i.e., far away from the initiation of the thermoplastic instability, as expected from Fig. 3(b) in the main text.
\vspace{5mm}
\begin{figure}[ht]
\centering
\includegraphics[width=0.7 \linewidth]{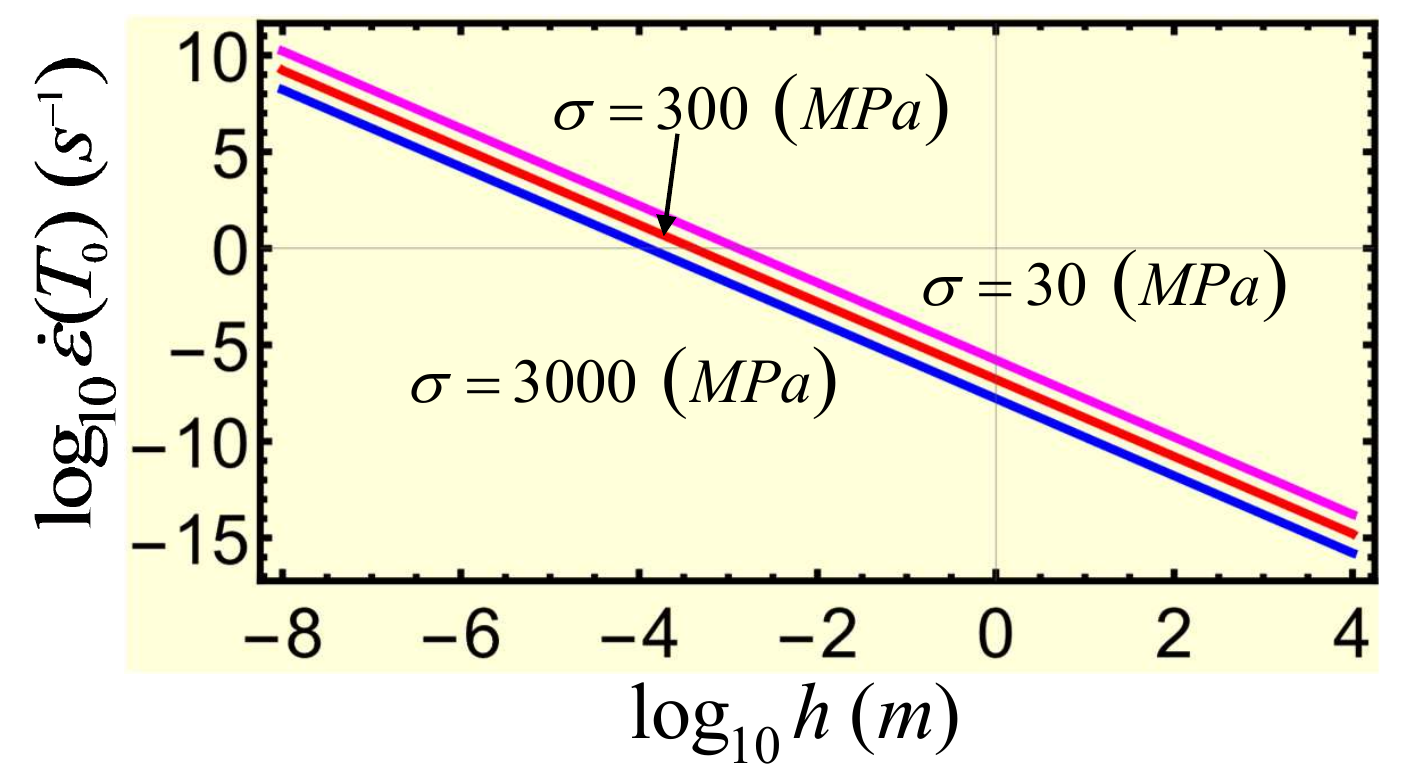}
\caption{{\bf Relationships between $\log_{10}\dot{\vep}(T_0) $ vs.  $\log_{10}h $ for three different stresses $\sg$.}}
\label{Fig-eps-h}
\end{figure}
\vspace{5mm}

It is convenient to present instability condition $B=B_c$ in the form
\bey
    \log_{10}(0.25/ k) +2  \log_{10} h +  \log_{10} \sg+  \log_{10} \dot{\vep}(T_0)= \log_{10}B_c,
\label{earth-16s}
\eey
see Fig. \ref{Fig-eps-h}. For the above parameters and $\dot{\vep}(T_0)=10^{-14}$ and  $10^{-12}\, s^{-1}$, the instability conditions can be satisfied for $h=4075 \, m$ and $h=407.5 \, m$, respectively. These parameters are in the range obtained in \cite{Ogava-87} numerically using linear perturbation analysis. Here, simple analytical expressions are derived. For the laboratory experiment on  for Mg$_2$GeO$_4$ from \cite{Riggs-Green-pileup-05},
$\dot{\vep}_0=2 \times 10^{-4}\, s^{-1}$ and  $\sg=1589\, MPa$, and the instability condition can be met for
$h=12.52 \, mm$, which is still very large for the laboratory experiment.

Does this mean that shear-induced melting is impossible in Earth and the laboratory due to very small observed band thickness?
Actually not. Let us assume that the initial shear band thickness can be much larger. This does not contradict to much smaller observed
thickness after phase transformation (including melting) because PT leads to further softening and, due to heterogeneities, may occur in a very narrow part of the initial
band. That is why after solid-solid PT or chemical reaction thickness of a band strongly decreases, and instability temperature cannot be reached at the laboratory scale
(but can be met in nature). If PT or reaction does not occur below the melting temperature at a high strain rate, melting can be reached.
Thus, taking $\dot{\vep}_0= 10\, s^{-1}$,  $\sg=1000\, MPa$, we obtain critical $h=70\, \mu m$, which may be achievable in large-volume high pressure apparatuses.
Taking $\sg=300\, MPa$ and  $h=0.07\, m$, which may be reasonable for a shear band in nature, we obtain critical
 $\dot{\vep}_0=3.33 \times 10\, s^{-5}$, which is not clear how to reach starting with  $10^{-14}-10^{-12}\, s^{-1}$
after all softening mechanisms unrelated to a PT .

\section{Phase equilibrium pressure-temperature diagram for olivine, $\beta$-spinel, and $\gamma$-spinel}

\begin{figure}[ht]
\centering
\includegraphics[width=0.6 \linewidth]{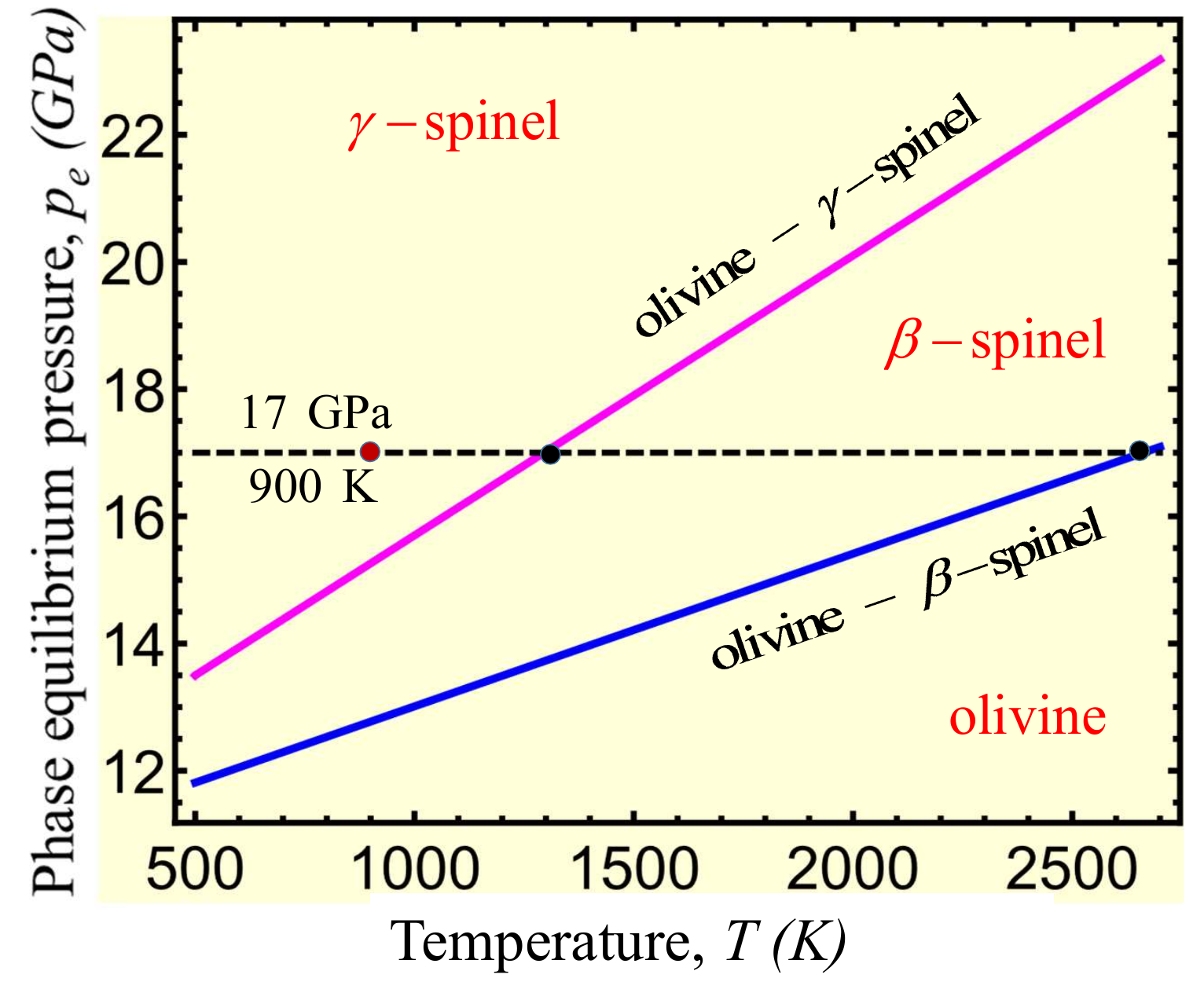}
\caption{{\bf Phase equilibrium pressure-temperature diagram for olivine, $\beta$-spinel, and $\gamma$-spinel based on \cite{Ottonello-etal-09}.} Red dot designates experimental parameters for the metastable olivine, black dots correspond to the phase equilibrium temperatures at 17 GPa.}
\label{Fig-phase diagram}
\end{figure}
\vspace{5mm}

 We assume that the minimum pressure for the direct PT, $  p_\vep^d (T)$, 
grows with temperature similar to the phase equilibrium pressure $p_e (T)$, see Fig. \ref{Fig-phase diagram}.
The equilibrium pressure olivine-$\beta$ spinel is $p_e (GPa)=14.5 + 0.0024 (T-1620)$ (approximated from \cite{Ottonello-etal-09}).  Then at $p=17\, GPa$ we have $T_e=2662\, K$, well above $T_s$ in Fig. 3 in the main text.
The phase equilibrium pressure olivine-$\gamma$ spinel is $p_e (GPa)=19.0 + 0.0044 (T - 1750)$  \cite{Ottonello-etal-09} and
at $p=17\, GPa$ we have $T_e=1295\, K$.
Since strain-induced PT
may occur even much below the phase equilibrium pressure,  we assume that PT olivine-$\gamma$ spinel occurs below $T_s=1695 K$,
the unstable stationary  temperature for straining   without PT for  $\dot{\vep}(T_0)=10^{-13}\, s^{-1}$ (Fig. 3 in the main text). Alternatively, PT olivine-$\beta$ spinel can be considered above 1295 K with slightly smaller $\vep_0$, which does not change conclusions. Moreover, if $\gamma$ spinel that appeared below 1295 K transforms back to the $\beta$ spinel, the same TRIP will take place like for $\beta \rightarrow \gamma$ spinel PT, since it is determined by $|\vep_0|$.

\section{Analysis of the lack of seismic activity below 660 km}

Lack of any of  the processes shown in Fig. 2 of the main text due to not meeting the required conditions  may explain lack of seismic activity below 660 km, where endothermic
 and slow disproportionation reaction from ringwoodite  to MgSiO$_3$ (bridgmanite)+ (Mg$_x$ Fe$_{1-x}$)O (magnesiowüstite) occurs.
 It is difficult to say which exactly process is missing because a counterargument may override each argument.
For example, one may say that reaction, in contrast to martensitic PT, requires a diffusive mass transport, and both nucleation and growth cannot be as fast as martensitic PT, which is proved for the proxy reaction albite$\rightarrow$jadeite + coesite \cite{Gleason-Green-09,Green-17}.
However, this may be true or not because large plastic shears strongly accelerate mass transport and chemical reaction as well \cite{zharov-1994,koch-1993,levitas+nesterenko+meyers-1998,takacs-2002,Takacs-2013,Balazetal-2013}, and it is unknown how do  shears affect this specific reaction. In particular, at friction surfaces the decomposition reaction of dolomite MgCa(CO$_3$)$_2 \rightarrow $MgO+CaO+2CO$_2$ completes within 0.006 s \cite{Green-etal-NatGeo-15} with temperature increase exceeding 1000 K.

The most probable reasons are:

(a) lack of initial shear localization in nanograined spinel before reaction due to grain sliding deformation
without orientational softening (which reduces $\vep (T_0)$ by a factor of $47$) and reduced dislocation activity,  which makes the transition to strain-induced PT and self-blown-up deformation-transformation-heating process impossible;

(b)  the higher initial temperature at 660 km (see \cite{Kawakatsu-11,Billen-20} and Fig. 1a in the main text); e.g., increase in $T_0$ from 900 K to 1000 K reduces parameter $M$ in Eq.(1) in the main text by a factor of 653, and

(c) low initial strain rate below 660 km  \cite{Billen-20} reduces the final strain rate proportionally.

One of the conditions for PT-induced instability mentioned in \cite{Green-07,Green-17} is the exothermic character of the olivine-spinel PT, leading to runaway heating. At the same time, the reaction from ringwoodite to bridgmanite + magnesiowüstite is endothermic and cannot produce instability and  earthquakes below 600 km. However, for coupled strain-induced PT-TRIP process, plastic heating during PT and contribution of PT heat ($100 K$ \cite{Sung-Burns-76}) in temperature increase from 900 to $T_s=1400-1800 \,K$ is small.
Thus, we do not think that the exothermic character of PT alone  is critical. In laboratory experiments, temperature change within the shear band is negligible.

Exothermic PT was utilized in \cite{green-etal-15} also to explain nanograined spinel structure.
The temperature increase due to PT heat increases the driving force for PT and causes runaway nucleation under growth-inhibited
conditions.
Suppose a slight temperature increase would be the reason for a drastic increase in nucleation rate. In that case, runaway nucleation should occur
everywhere rather than to localize within anticracks, especially in  hotter regions of the metastable  olivine slab  closer
 to its boundary with spinel. It is also unclear why growth is slow at such a large thermodynamic driving force that causes runaway nucleation. At the same time, nucleation at dislocations and dislocations pileups leads to nanograined structure
because of growth arrest due to a strong reduction of stresses away from the defect tip \cite{levitas-mechchem-04,levitas-prb-04,Levitas-JPCM-18,Levitas-MT-19}.

\section{Relation to some previous works}

 TRIP is well known to the geological community, but it was considered to having a small effect \cite{Frohlich-06,Kirby-87,Karato-etal-01,Poirier-00}.
This is correct in general, but for a properly oriented shear band where $\tau \rightarrow  \tau_{y}$, plastic shear tends to infinity
(see Eq.(\ref{earth-14s-a}) and Fig. 4(a) in the main text). Shear banding and TRIP  are observed in DAC
experiments  in  fullerene \cite{Kulnitskiyetal-16} and BN \cite{Levitasetal-JCP-BN-06}  despite the PTs to stronger high-pressure phases. For PT from hexagonal to superhard wurtzitic BN, TRIP was evaluated to be 20 times larger than prescribed shear \cite{Levitasetal-JCP-BN-06}. Shear banding during PT is possible if the yield strength $\tau_y$ during PT does not increase despite the high strain rate and strength of the high-pressure phase, which supports our conservative hypothesis  $\tau_y=const$.
Positive feedback between PT and TRIP without heating was suggested in  \cite{Levitasetal-JCP-BN-06} but without any equations.
Reaction-induced plasticity (RIP) similar to TRIP was revealed for a chemical reaction within a 
shear band in {\sl Ti-Si}  powder mixture
\cite{levitas+nesterenko+meyers-1998}, and TRIP-induced adiabatic heating was considered as a factor promoting reaction rate.
However, mechanochemical feedback was not claimed since kinetics was considered within the theory for stress-induced reactions instead of strain-induced.

 It is shown in \cite{Markenscoff-21} based on the elegant dynamic solution for ‘‘pancake-like’’ flattened ellipsoidal Eshelby inclusion that it can grow self-similarly above some critical pressure. It is also derived that in order for the total strain energy to be finite (and not zero) in the  inclusion with tending to zero thickness,  deviatoric eigen strain (without specification of its nature) must tend to infinity (even under hydrostatic compression), which "explains" deviatoric character of the deep-earthquake source. This argument is  unphysical: why should zero-thickness inclusion "desire" to have nonzero strain energy? Eigen strain in inclusion should be determined by processes in inclusion, like PT and plasticity, which is done in the current paper. Huge TRIP shear in Eq.(\ref{earth-14s-a}) after complete PT explains deviatoric character of the deep-earthquake source.  Also, plasticity (that significantly affects the stress-strain fields, reduces thermodynamic driving force, and may arrest PT  \cite{Levitas-etal-PhilMag-02}) is neglected in \cite{Markenscoff-21}, as well as interfacial energy.

\newpage

\renewcommand{\thepage}{\arabic{page}}

\baselineskip14pt
\belowdisplayskip10pt
\belowdisplayshortskip10pt

\end{document}